\documentclass[twocolumn]{aastex631}

\begin{document}

\title{ADF22-WEB: ALMA and JWST (sub)kpc-scale views of dusty star-forming galaxies in a $z\approx$3 proto-cluster}

\author{Hideki Umehata}
\affiliation{Institute for Advanced Research, Nagoya University, Furocho, Chikusa, Nagoya 464-8602, Japan}
\affiliation{Department of Physics, Graduate School of Science, Nagoya University, Furocho, Chikusa, Nagoya 464-8602, Japan}
\author{Mariko Kubo}
\affiliation{Astronomical Institute, Tohoku University, 6-3, Aramaki, Aoba, Sendai, Miyagi, 980-8578, Japan}
\author{Ian Smail}
\affiliation{Centre for Extragalactic Astronomy, Department of Physics, Durham University, South Road, Durham DH1 3LE, UK}
\author{Bret D.~Lehmer}
\affiliation{Department of Physics, University of Arkansas, 226 Physics Building, 825 West Dickson Street, Fayetteville, AR 72701, USA}
\author{Erik~B.~Monson}
\affiliation{Department of Astronomy and Astrophysics, The Pennsylvania State University, 525 Davey Lab, University Park, PA 16802, USA}
\author{Kouichiro~Nakanishi}
\affiliation{National Astronomical Observatory of Japan, 2-21-1 Osawa, Mitaka, Tokyo 181-8588, Japan}
\affiliation{Department of Astronomical Science, The Graduate University for Advanced Studies, SOKENDAI, 2-21-1 Osawa, Mitaka, Tokyo
181-8588, Japan}
\author{Yuichi~Matsuda}
\affiliation{National
Astronomical Observatory of Japan, 2-21-1 Osawa, Mitaka, Tokyo 181-8588, Japan}
\affiliation{Department of Astronomical Science, The Graduate University for Advanced Studies, SOKENDAI, 2-21-1 Osawa, Mitaka, Tokyo
181-8588, Japan}



\begin{abstract}
We present a morphological analysis of ALMA and JWST NIRCam images of nine dusty star-forming galaxies (DSFGs) at $z_{\rm spec} \approx 3.09$, all embedded within the cosmic web filaments at the SSA22 proto-cluster core. The ALMA 870\,$\mu$m and 1.1\,mm images are obtained at spatial resolutions ranging from 0.5$^{\prime\prime}$ to 0.05$^{\prime\prime}$ (350\,pc at $z=3.09$). The high-resolution images enable us to resolve inner structures traced by dust continuum, identifying compact dusty cores, clumps, and offset ridges within bars. S\'ersic profile fit was performed for both ALMA 870\,$\mu$m and NIRCam F444W images at comparable resolutions ($\sim0.15^{\prime\prime}$).
The S\'ersic index measured for 870\,$\mu$m, masking bright regions, indicates values close to unity, suggesting that dust emission arises from disks with superimposed compact core components. For the JWST F444W images (restframe $\sim1 \mu$m), the S\'ersic indices range between $n_{\rm F444W} \sim 1-3$, pointing to the coexistence of bulges and stellar disks in these DSFGs. 
A comparison of dust mass surface density, $n_{\rm F444W}$, and F200W--F444W color (restframe $\sim0.5-1 \mu$m) reveals diversity among the DSFGs, likely reflecting different evolutionary stages including some DSFGs with red cores, indicating ongoing rapid bulge growth phases heavily obscured by dust. The predominantly disk-like morphologies observed in most DSFGs in the proto-cluster core contrast sharply with early-type morphologies that dominate the highest density environment in the local universe. This suggests that we are witnessing the early formation of the morphology-density relation, as massive galaxies undergo rapid growth as late-type galaxies fueled by cosmic web gas filaments.
\end{abstract}

\keywords{galaxies:  evolution - galaxies:  star formation}


\section{Introduction} \label{sec:intro}

The cold dark matter (CDM) model predicts that galaxy formation and evolution are tightly linked to the growth of large-scale cosmic structures, known as the cosmic web, which consists of interconnected dark matter and baryonic filaments shaping galaxy distribution (e.g., \citealt{1996Natur.380..603B}; \citealt{1999MNRAS.302..111L}; \citealt{2005Natur.435..629S}). Understanding the environments in which galaxies reside, including their positions within the cosmic web, is therefore essential for unraveling the processes driving galaxy formation and evolution across cosmic time.

In the local universe, dense environments such as galaxy clusters are predominantly populated by passive, early-type galaxies, whereas less dense regions are characterized by star-forming, late-type galaxies (e.g., \citealt{1980ApJ...236..351D}). Similarly, in the local-to-nearby universe, clusters exhibit a prominent red sequence of passive galaxies (e.g., \citealt{1997ApJ...483..582E}; \citealt{1998A&A...334...99K}; \citeyear{2007MNRAS.377.1717K}; \citealt{2012MNRAS.426.2994S}). The passive nature of these galaxies indicates that the majority of their stars formed at least $\sim1-2$ Gyr prior to the epoch of observation. Cosmological simulations also predict that the progenitors of massive elliptical galaxies, which are in actively star-forming phases, are preferentially located in high-density environments at $z\gtrsim2-3$ (e.g. \citealt{2006MNRAS.366..499D}).

Dusty star-forming galaxies (DSFGs), also known as submillimeter galaxies (SMGs), are characterized by their remarkable submillimeter flux resulting from intense, dust-obscured star formation (for reviews, see \citealt{2002PhR...369..111B}; \citealt{2014PhR...541...45C}; \citealt{2020RSOS....700556H}). Since the discovery of DSFGs (\citealt{1997ApJ...490L...5S}; \citealt{1998Natur.394..241H}; \citealt{1998Natur.394..248B} \citealt{1998MNRAS.298..583I}), their starburst nature in the early universe has led to suggestions that they are plausible progenitors of local elliptical galaxies (e.g., \citealt{1999ApJ...515..518E}; \citealt{2014ApJ...782...68T}). Testing this scenario requires uncovering the environments where DSFGs reside, characterizing the stellar assembly occurring within DSFGs, and connecting the formation and evolution of DSFGs to the larger structure formation processes.

Before the advent of the Atacama Large Millimeter/submillimeter Array (ALMA), substantial efforts were made to study DSFGs and their environments (e.g. \citealt{2003Natur.425..264S}; \citealt{2004ApJ...611..725B}; \citealt{2005MNRAS.359.1165G}; \citealt{2009ApJ...694.1517D}; \citealt{2009Natur.459...61T}; \citealt{2012Natur.486..233W}; \citealt{2014MNRAS.440.3462U}). A significant breakthrough came when \cite{2015ApJ...815L...8U} identified eight individual DSFGs with spectroscopically confirmed redshifts ($z_{\rm spec}$) within a $z=3.1$ proto-cluster core. This marked the first discovery of a large number of DSFGs associated with proto-clusters, free from source confusion at submillimeter and millimeter wavelengths. The advent of ALMA has since expanded our knowledge of DSFGs in proto-clusters, reaching redshifts as high as $z \gtrsim 5$ (e.g., \citealt{2013Natur.496..329R}; \citealt{2018ApJ...856...72O}; \citealt{2018Natur.556..469M}; \citealt{2024ApJ...961...69S}).

Despite significant advancements, one major obstacle in understanding the nature, origins, and evolution of DSFGs is the lack of spatial resolution. The heavy dust extinction in DSFGs obscures a substantial fraction of stellar emission in wavelengths accessible with the Hubble Space Telescope (HST) (e.g. \citealt{2010MNRAS.405..234S}; \citealt{2015ApJ...799..194C}; \citealt{2021ApJ...919...51M}). Meanwhile, the limited spatial resolution of the Spitzer Space Telescope has made it challenging to study stellar structures in the rest frame near- to mid-infrared wavelengths, which are better suited for tracing stellar mass distributions (e.g. \citealt{2011ApJ...740...96H}; \citealt{2011MNRAS.415.1479W}; \citealt{2014ApJ...788..125S}; \citealt{2014MNRAS.440.3462U}). 

In terms of dust continuum observations, while ALMA has significantly improved the identification of individual DSFGs through snapshot surveys and deep mosaics (e.g. \citealt{2013ApJ...768...91H}; \citealt{2013ApJ...767...88W}; \citealt{2015ApJ...810..133I}; \citealt{2017MNRAS.466..861D}; \citealt{2019MNRAS.487.4648S}), achieving high spatial resolution data with high fidelity remains a challenge even with ALMA. These limitations continue to hinder a comprehensive understanding of DSFGs, both in general fields and in biased environments, such as proto-clusters.

We are gradually overcoming these difficulties. For instance, high-resolution mapping of the dust continuum in DSFGs at sub-kpc scales has unveiled internal structures indicative of bars and spiral arms (e.g., \citealt{2016ApJ...829L..10I}; \citealt{2016ApJ...833..103H}; \citeyear{2019ApJ...876..130H}; \citeyear{2025ApJ...978..165H}; \citealt{2018Natur.560..613T}; \citealt{2018ApJ...859...12G}). Furthermore, the James Webb Space Telescope (JWST) has dramatically improved our ability to resolve stellar structures. Since its launch, studies have investigated the stellar structures in DSFGs, with some reporting disc-like profiles indicative of secular evolutionary processes, while others reveal signatures of interactions (e.g., \citealt{2022ApJ...939L...7C}; \citealt{2023ApJ...942L..19C}; \citealt{2023A&A...673L...6C}; \citealt{2023A&A...676A..26G}; \citeyear{2024A&A...691A.299G}; \citealt{2025ApJ...978..165H}).

Despite these advances, critical challenges remain. First, it is necessary to construct samples observed with both ALMA and JWST at comparably high resolutions to enable a comprehensive interpretation of DSFGs (e.g. \citealt{2025ApJ...978..165H}). Second, as noted earlier, it is fundamentally important to connect DSFGs to their underlying environments, requiring detailed studies in notable proto-clusters in the early universe.

Motivated by these challenges, we have initiated the ``ADF22-WEB'' project (\citealt{2024arXiv241022155U}). This is an extension of the ALMA Deep Field in SSA22 (ADF22) project, which has been developed since ALMA Cycle-2 (e.g., \citealt{2015ApJ...815L...8U}; \citeyear{2017ApJ...835...98U}; \citeyear{2018PASJ...70...65U}; \citeyear{2020A&A...640L...8U}; \citealt{2017PASJ...69...45H}). In the core region of the SSA22 proto-cluster at $z=3.1$ (\citealt{1998ApJ...492..428S}; \citealt{2004AJ....128.2073H}; \citealt{2005ApJ...634L.125M}), about twenty DSFGs and eight X-ray AGNs, including six DSFGs hosting X-ray AGNs, are embedded in Ly\,$\alpha$ filaments stretching over 4 comoving Mpc (\citealt{2019Sci...366...97U}; S. Huang et al., in preparation). This field provides an invaluable laboratory for investigating and uncovering how galaxies form and evolve along cosmic web filaments.

In this paper, we present new ALMA and JWST observation of nine DSFGs at $z_{\rm spec}=3.09$ in the ADF22 field and discuss their nature based on resolved views of stellar and dust structures. The structure of this paper is as follows. In \S 2, we describe the observations with ALMA and JWST. In \S 3, we present a detailed morphological analysis of the ALMA continuum images. In \S 4, we present the JWST/NIRCam images and compare them with the ALMA images. In \S 5, we discuss the origins of the dust emission, bulge formation, morphological transformation, and the influence of the cosmic web on galaxy evolution. We conclude in \S 6.
We adopt a standard concordance cosmology with $H_0=70$\,km\,s$^{-1}$\,Mpc$^{-1}$, $\Omega_{\rm m}=0.30$, and $\Omega_\Lambda=0.70$. Here, $H_0$ is the Hubble constant, and $\Omega_{\rm m}$ and $\Omega_\Lambda$ are the matter density and dark energy density at the present time, respectively. This cosmology gives a scale of 7.63\,kpc per arcsec at $z=3.09$.

\section{Observation and data reduction} \label{sec:obs}

\subsection{Target Selection} \label{sec:obs_target}

Among the 16 DSFGs at $z_{\rm spec}\approx3.09$ in ADF22 reported in \cite{2019Sci...366...97U} (see also \citealt{2015ApJ...815L...8U}; \citeyear{2017ApJ...835...98U}), we selected the brightest six DSFGs (ADF22.A1, A3, A4, A5, A6, and A7) as primary targets. They are ULIRG-class DSFGs (IR luminosity $L_{\rm IR}\gtrsim10^{12}$\,L$_\odot$, \citealt{2017ApJ...835...98U}) and suitable to investigate the nature of such {\it classical}, relatively bright DSFGs. Four of them have been known to harbor X-ray AGNs (\citealt{2010ApJ...724.1270T}; \citealt{2015ApJ...815L...8U}; \citeyear{2019Sci...366...97U}; \citealt{2023ApJ...951...15M}). We also discuss three additional DSFGs (ADF22.A10, A11, and A16) as secondary targets. These DSFGs are located in the vicinity of the primary target ADF22.A4 and were simultaneously observed and detected by both ALMA and JWST, while the fidelity of the dust continuum emission is reduced slightly due to their relatively submm-faint nature.   

\subsection{ALMA Observation}
\label{sec:obs_alma}

\begin{deluxetable*}{cccccc}
\tabletypesize{\scriptsize}
\tablewidth{0pt} 
\tablecaption{ALMA band 6 and 7 observation parameters \label{tab:ALMA_obs}}
\tablehead{
\colhead{ALMA Band} & \colhead{Wavelength} & \colhead{Beam size}  & \colhead{rms Noise} & \colhead{Beam size}  & \colhead{rms Noise} \\
 \colhead{} &   \colhead{}&  \colhead{} & \colhead{($\mu$Jy\,beam$^{-1}$)}   &  \colhead{} & \colhead{($\mu$Jy\,beam$^{-1}$)}  
} 
\startdata 
Band-6 & 1.1\,mm & $0.047^{\prime\prime}\times0.039^{\prime\prime}$ & 17 & $0.079^{\prime\prime}\times0.072^{\prime\prime}$ & 19 \\
Band-7 & 870\,$\mu$m & $0.15^{\prime\prime}\times0.14^{\prime\prime}$ & 28 (24) & $0.54^{\prime\prime}\times0.48^{\prime\prime}$ & 53 (47) \\
\enddata
\tablecomments{A mosaic image was created by combining the two fields, ADF22.A1 and ADF22.A7, resulting in a slightly improved noise level for the two DSFGs in Band-7 (as indicated in the brackets). The RMS noise levels at the source positions were measured after applying the primary beam correction.}
\end{deluxetable*}

\subsubsection{Band-6 observations}

Observations in ALMA Band-6 to map the 1.1 mm dust continuum were conducted in August 2021 as a Cycle-7 program (PI Umehata, 2019.1.00008.S). The data were taken in four execution blocks (EBs). For all, the array was configured in C43--8 with 43--49 antennas, which offered baseline lengths of 46\,m to 12600\,m. The recipitable water vapor (PWV) $\sim0.5$\,mm is suitable for Band-6 observations. The six primary target fields were observed and a total on-source time for each source was 31.4\,min. J2253+1608 was used as a bandpass and flux calibrator, while J2226+0052 was regularly checked for phase calibration.
The four spectral windows were centered at frequencies of 253.0, 254.8, 267.0, and 269.0 GHz. The set-up encompasses CO(9--8) and OH$^{+}(1_1 \rightarrow 0_1)$ lines. We masked the lines if they are detected. 
Data reduction was carried out using version 6.1.0 and 6.5.0 of the {\sc casa} package. The continuum image was generated from line-free channels using the {\sc tclean} task. The representative frequency is 261.51\.GHz (or 1.146\,mm).
We create a mosaic by combining two nearby fields, ADF22.A1 and ADF22.A7, to enhance sensitivity.
We employ the \texttt{usemask} parameter set to \texttt{auto-multithresh}. In the initial pass of mask creation for each channel, only pixels with a signal-to-noise ratio (S/N) of 4.5 or higher are included (\texttt{noisethreshold} = 4.5). Subsequently, the mask is expanded to encompass regions with lower S/N values (\texttt{lownoisethreshold} = 1.5). The regions identified through automasking are then cleaned down to the 2$\sigma$ level (\texttt{nsigma} = 2 in \texttt{TCLEAN}).
Images at two different spatial resolutions were obtained to track dust continuum emission at a different spatial scale. 
Using the Briggs weighting with the robust parameter 1.0, images have angular resolution of $0.05^{\prime\prime}$. Applying natural weighting with {\sc uvtaper} set to $0.038^{\prime\prime}$ produced a synthesized beam size of $0.08^{\prime\prime}$. We summarize the beam size and rms noise level in Table \ref{tab:ALMA_obs}.

\subsubsection{Band-7 observations}

ALMA Band-7 observations for the six fields centered on the brightest DSFGs were performed as a Cycle-8 project (PI Umehata, 2021.1.00071.S). The C43-3 and C43-6 array configurations were utilized in May and July in 2022. The combination of array configurations are designed to be sensitive to emission on various scales. Data were obtained through two EBs in C43-3 and three EBs in C43-6, yielding on-source time per field of 8\,min and 21\,min, respectively. The covered base line lengths by the two array configurations ranges from 15\,m to 2617\,m, and 42--27 12-m antennas were used. PWV was $\sim0.5$\,mm.
The correlator was set up with two spectral windows
of 1.875\,GHz bandwidth (dual polarization) each per sideband. The spectral windows had central frequencies of 336.5, 338.4, 348.5, and 350.5\,GHz, respectively. The representative frequency is 343.47\,GHz, which corresponds to $\approx870$\,$\mu$m at the observed frame.
Data reduction was performed using version {\sc casa} 6.5.0, mapped with the {\sc tclean} task as in the case of Band-6 data. Similarly we combined the ADF22.A1 and ADF22.A7 fields to obtain a single image. We make two-types of images. First we adopted the Briggs weighting with the robust parameter 0.5, which resulted in $\sim0.15^{\prime\prime}$ resolution. Second we imaged only with the C43-3 array data with natural weighting to investigate the flux distribution in a way which is not well resolved. These maps have $\sim0.5^{\prime\prime}$ resolutions. Observational parameters are summarized in Table \ref{tab:ALMA_obs}.

\subsection{JWST Observation}
\label{sec:obs_jwst}

The ADF22 field was observed with JWST/NIRCam as part of the JWST Cycle-2 program (PI Umehata, GO 3547) as reported in \cite{2024arXiv241022155U}. Here we briefly describe the observation. Images are obtained with four fileters; F115W, F200W, F356W, and F444W. 
Each filter had an exposure time of 1869\,sec using the STANDARD subpixel dither pattern without primary dithering. The observations employed the MEDIUM8 readout mode. Data reduction followed the JWST calibration pipeline (v.1.8.4) with additional steps for ``snowball'' removal, wisp subtraction, and 1/f noise correction, based on CEERS team guidelines \citep{2023ApJ...946L..12B}. The 5$\sigma$ limiting magnitudes for a 0.2$^{\prime\prime}$ diameter aperture are 29.74, 29.01, 29.15, 29.49, 29.00 AB mag for F115W, F200W, F356W, and F444W, respectively, with the PSF of each image matched to F444W.
To register coordinates accurately and perform fair comparison between JWST and ALMA images, we first match the xy coordinates among the four-band NIRCam images to each other, and reconfigure the wcs based on the Gaia DR3 stars’ coordinates. We then refined the alignment by minimizing systematic offsets between the emission peaks in the ALMA 870\,$\mu$m and NIRCam F444W images, assuming that there is no systematic offset between the two (though each peak does not necessarily match perfectly). This approach, utilizing the advantages of multiple detections in the same field (i.e., ADF22), ensures consistency of astrometry between ALMA and NIRCam images within the native pixel scale of the F444W image (0.063$^{\prime\prime}$\,pix$^{-1}$).

\section{Measurements on dust continuum} \label{sec:dust}

\subsection{Overview and flux measurements} \label{sec:dust_flux}

\begin{figure*}
\epsscale{0.87}
\plotone{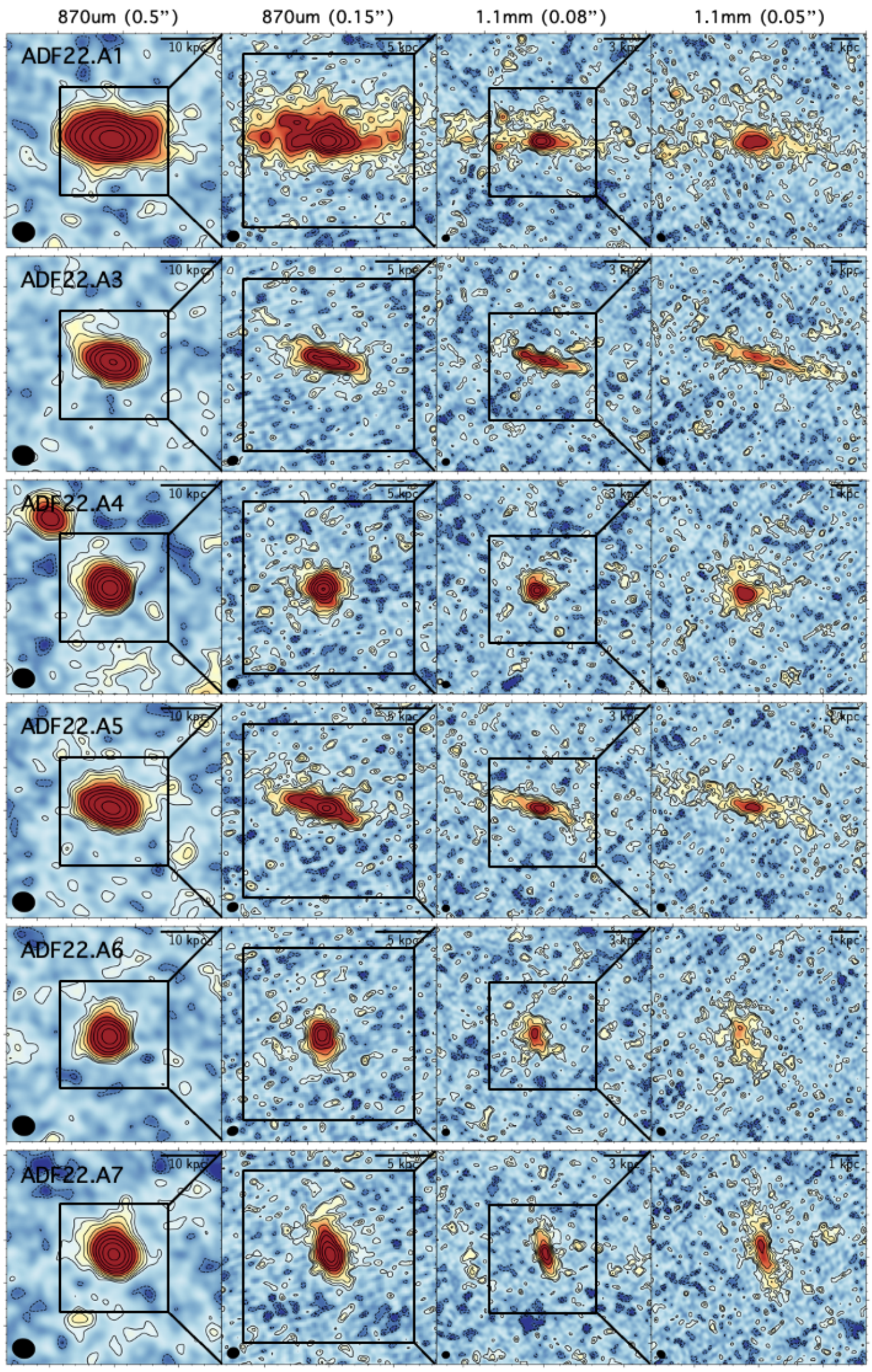}
\caption{
The cutout images of the six brightest $z=3.09$ DSFGs in ADF22. From left to right, close-up views are shown, with each inserted square indicating the area highlighted in the adjacent tile. Each map presents 870\,$\mu$m images with resolutions of $0.5^{\prime\prime}$ and $0.15^{\prime\prime}$, as well as 1.1\,mm images with resolutions of $0.08^{\prime\prime}$ and $0.05^{\prime\prime}$, respectively. The panel sizes are $5^{\prime\prime} \times 5^{\prime\prime}$, $2.5^{\prime\prime} \times 2.5^{\prime\prime}$, $2^{\prime\prime} \times 2^{\prime\prime}$, and $1^{\prime\prime} \times 1^{\prime\prime}$, respectively.
Contours correspond to $\pm \sigma \times 1.5^n$, with $n=1$ to $n=11$.
The dust emission is resolved down to 0.05$^{\prime\prime}$ (350\,pc at $z=3.09$), revealing substructures that suggest bright central cores and offset ridges in bars. We discuss the origins of the dust emission in \S \ref{sec:discussion_origin}.
}
\label{fig:almamaps}
\end{figure*}

\begin{figure*}
\epsscale{0.87}
\plotone{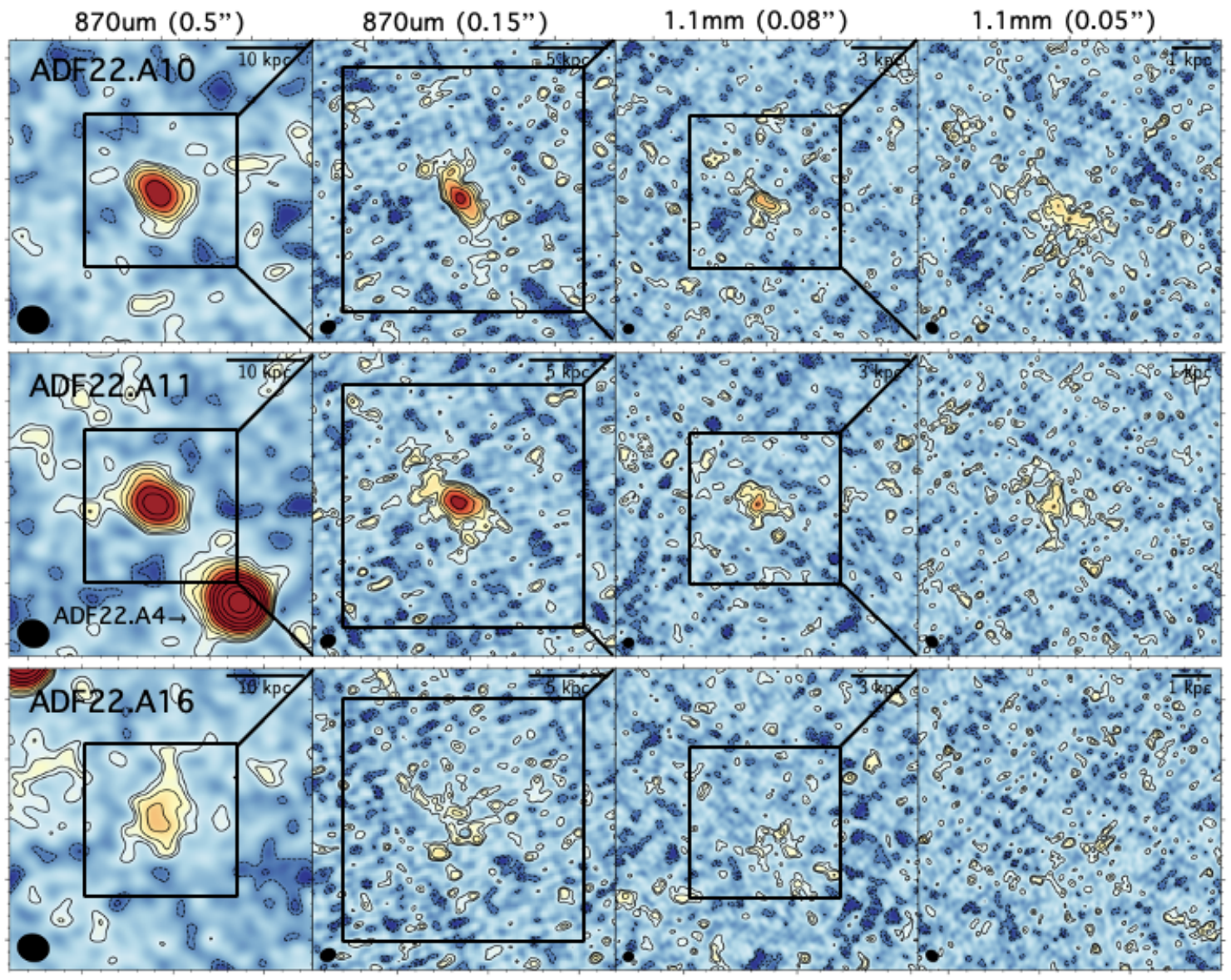}
\caption{
The cutout images of three $z=3.09$ DSFGs in the vicinity of ADF22.A4. Panels are as shown in Fig.~\ref{fig:almamaps}.
}
\label{fig:almamaps_faint}
\end{figure*}

\begin{deluxetable*}{lcccccccc}
\tabletypesize{\scriptsize}
\tablewidth{0pt} 
\tablecaption{ALMA Flux measurements \label{tab:ALMA_flux}}
\tablehead{
\colhead{ID} & \colhead{R.A.} & \colhead{Dec.} & \colhead{$S_{\rm 870\mu m, 0.5^{\prime\prime}}$}  & \colhead{$S_{\rm 870\mu m, 0.15^{\prime\prime}}$} & \colhead{$S_{\rm 1.1mm, 1.0^{\prime\prime}}$} & \colhead{$S_{\rm 1.1mm, 0.08^{\prime\prime}}$}  & \colhead{$S_{\rm 1.1mm, 0.05^{\prime\prime}}$} & \colhead{$S_{\rm 1.1mm, 0.05^{\prime\prime}}^{\rm peak}$} \\
\colhead{}  &  \colhead{IRCS} &  \colhead{IRCS} &  \colhead{(mJy)} & \colhead{(mJy)}   &  \colhead{(mJy)} &  \colhead{(mJy)} & \colhead{(mJy)}  &  \colhead{(mJy)}}
\startdata 
ADF22.A1  & 22:17:32.42 & 00:17:43.86 & $14.67 \pm 0.18$ & $14.40 \pm 0.19$ & $6.15 \pm 0.06$ & $5.86 \pm 0.15$ & $3.71 \pm 0.11$ & $0.26 \pm 0.02$\\
ADF22.A3  & 22:17:35.15 & 00:15:37.24 & $5.09 \pm 0.14$  & $4.75 \pm 0.13$  & $2.23 \pm 0.04$ & $2.28 \pm 0.08$ & $2.18 \pm 0.09$ & $0.19 \pm 0.02$\\
ADF22.A4  & 22:17:36.97 & 00:18:20.68 & $5.23 \pm 0.13$  & $4.98 \pm 0.12$  & $2.26 \pm 0.09$ & $2.20 \pm 0.08$ & $2.11 \pm 0.09$ & $0.28 \pm 0.02$\\
ADF22.A5  & 22:17:31.49 & 00:17:58.08 & $6.67 \pm 0.15$  & $6.08 \pm 0.14$  & $2.53 \pm 0.06$ & $2.40 \pm 0.09$ & $2.18 \pm 0.10$ & $0.27 \pm 0.02$\\
ADF22.A6  & 22:17:35.84 & 00:15:58.95 & $3.60 \pm 0.12$  & $3.20 \pm 0.10$  & $1.43 \pm 0.04$ & $1.63 \pm 0.08$ & $1.14 \pm 0.07$ & $0.14 \pm 0.02$\\
ADF22.A7  & 22:17:32.20 & 00:17:35.68 & $4.64 \pm 0.12$  & $4.17 \pm 0.10$  & $2.03 \pm 0.04$ & $1.78 \pm 0.07$ & $1.64 \pm 0.07$ & $0.22 \pm 0.02$\\
ADF22.A10 & 22:17:37.10 & 00:18:26.77 & $1.50 \pm 0.09$  & $1.68 \pm 0.08$  & $0.82 \pm 0.05$ & $0.46 \pm 0.04$ & $0.57 \pm 0.05$ & $0.11 \pm 0.02$\\
ADF22.A11 & 22:17:37.06 & 00:18:22.32 & $1.71 \pm 0.11$  & $1.67 \pm 0.10$  & $0.85 \pm 0.03$ & $0.60 \pm 0.06$ & $0.44 \pm 0.05$ & $0.09 \pm 0.02$\\
ADF22.A16 & 22:17:36.81 & 00:18:18.07 & $0.74 \pm 0.10$  & $0.46 \pm 0.07$  & $0.41 \pm 0.06$ & $0.06 \pm 0.02$ & $<0.06$ & $<0.06$\\
\enddata
\tablecomments{
Coordinates are measured with {\sc casa/imfit} using the 870\,$\mu$m maps at 0.5$^{\prime\prime}$. Flux in the 1.1\,$\mu$m maps at 1.0$^{\prime\prime}$ ($S_{\rm 1.1mm, 1.0^{\prime\prime}}$) are measured with {\sc casa/imfit}. For the remaining cases, fluxes enclosed with 2$\sigma$ contours are measured. We also report the 1.1\,$\mu$m peak flux density at 0.05$^{\prime\prime}$ ($S_{\rm 1.1mm, 0.05^{\prime\prime}}^{\rm peak}$).}
\end{deluxetable*}

Fig.~\ref{fig:almamaps} presents ALMA dust continuum maps of the six bright DSFGs at $z=3.09$, shown at various spatial resolutions and scales. Each resolution offers unique advantages. The 870\,$\mu$m maps, obtained at resolutions of 0.5$^{\prime\prime}$ (3.8\,kpc at $z=3.09$) and 0.15$^{\prime\prime}$ (1.1\,kpc at $z=3.09$), provide both (nearly) unresolved and resolved views of the galaxies on a whole galaxy scale with high significance. In contrast, the 1.1\,mm maps, with resolutions of 0.08$^{\prime\prime}$ (600\,pc at $z=3.09$) and 0.05$^{\prime\prime}$ (350\,pc at $z=3.09$), reveal detailed inner structures, though they less sensitive for tracing the total emission of the galaxies with high fidelity.
These high-resolution maps uncovered complex inner structures of the dust continuum emission within the proto-cluster DSFGs (Fig.~\ref{fig:almamaps}.)

A first step we carried out was to measure flux densities of the DSFGs in these maps. We use the 870\,$\mu$m images at $0.5^{\prime\prime}$ to estimate the total flux at the wavelength. We found that a Gaussian fit using {\sc casa/imfit} often leaves significant residuals (both positive and negative), suggesting that the use of a single Gaussian does not work to describe observed surface brightness profiles even at the 3.8\,kpc resolution. Here we measure fluxes, counting emissions enclosed by 2$\sigma$ contours (To measure profiles more accurately, we will perform a more detailed profile fit in \S \ref{sec:almafit}). 
These measurements are generally consistent with those of {\sc casa/imfit} ($\langle S_{\rm Imfit}/S_{\rm cntr}\rangle=1.03\pm0.07$). The ratio is calculated as the weighted average of the individual measurement ratios and the error is determined using the standard error of the weighted average. Considering the complex morphologies in finer resolution maps, we generally adopt this method and measure fluxes enclosed within 2$\sigma$ contours for each map. These measurements were summarized in Table~\ref{tab:ALMA_flux}. Peak fluxes in the 1.1mm images at 0.05$^{\prime\prime}$ were also listed.
We also added the 1.1\,mm total fluxes at 1.0$^{\prime\prime}$ resolution measured with the {\sc casa/imfit} (The image were obtained using independent Band-6 observations with the modest angular resolution. Details will be presented in S. Huang et al. in preparation). 

\subsection{Recovered flux ratios}

\begin{figure}
\epsscale{1.16}
\plotone{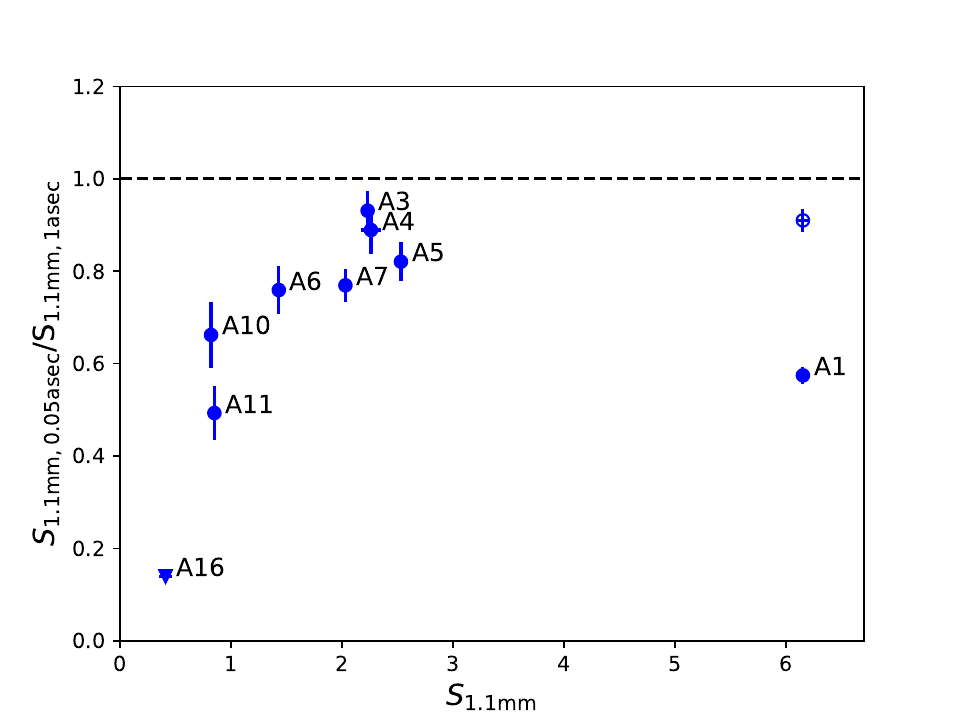}
\caption{
The 1.1\,mm flux ratio between two angular resolutions ($S_{\rm 1.1mm, 0.05asec}$/$S_{\rm 1.1mm, 1asec}$) is shown as a function of the observed total 1.1,mm flux $S_{\rm 1.1mm, 1asec}$. The open circle represents the case of ADF22.A1 at 0.08$^{\prime\prime}$. The fraction of recovered flux at 0.05$^{\prime\prime}$ is broadly correlated with the total flux. This correlation may reflect differences in the intrinsic nature of dusty star-forming activity between bright and faint DSFGs, although we note several caveats.
}
\label{fig:flux_recov}
\end{figure}

The 870\,$\mu$m flux measured at $0.15^{\prime\prime}$ resolution and the 1.1\,mm flux at $0.08^{\prime\prime}$ resolution account for approximately $90\%-110\%$ of the total fluxes measured at $0.5^{\prime\prime}-1.0^{\prime\prime}$ for bright DSFGs (Table~\ref{tab:ALMA_flux}). In the case of the $0.08^{\prime\prime}$ resolution 1.1\,mm images, the maximum recoverable scale is $\approx0.5-0.6^{\prime\prime}$ and the spatial resolution corresponds to scales of 600\,pc at $z=3.1$. The high fraction of recovery fluxes suggests that the majority of the flux is recovered at these scales. 
On the other hand, the diversity of flux recovery rates becomes more apparent as the resolution further increases and/or the total flux decreases. 

Fig.~\ref{fig:flux_recov} shows the 1.1\,mm flux ratio between two angular resolutions ($S_{\rm 1.1mm, 0.05asec}$/$S_{\rm 1.1mm, 1asec}$), which represents the fraction of recovered flux at very high spatial resolution, as a function of the observed total 1.1\,mm flux for each galaxy. Since the image at 1.0$^{\prime\prime}$ has a slightly lower observed frequency (1.167\,mm) compared to that of the high-resolution image (1.146\,mm), we applied a correction factor (5\%) to $S_{\rm 1.1mm, 1asec}$ in the calculation of the fraction, assuming modified blackbody radiation with $T_{\rm d} = 30$\,K and $\beta = 2.0$.

One notable feature in the figure is the relatively low recovered fraction for ADF22.A1 compared to other bright DSFGs (ADF22.A3--ADF22.A7). This may be related to its exceptionally large size. As summarized in Table~\ref{tab:ALMA_profile}, the dust size of ADF22.A1 is $\gtrsim 2$ times larger than that of the other DSFGs. Assuming that both Band-6 and Band-7 data trace the same structure, its effective radius $R_{\rm e} \approx 0.62^{\prime\prime}$ is comparable to the maximum recoverable scale of the C43-8 observation, suggesting extended flux is being resolved out.

On the other hand, there appears to be a correlation between the total flux and the recovered fraction for the remaining DSFGs. Fainter sources tend to show lower recovery rates. One possible explanation is a difference in the mechanisms powering the dust continuum emission: brighter SMGs may exhibit more compact emission structures, while fainter systems may be more spatially extended. A caveat to this scenario is that fainter DSFGs inherently have lower signal-to-noise ratios, which could result in a lower fraction of significantly detected flux at a given threshold in the high-resolution image. Deeper imaging would help to better understand the underlying situation.

Overall, these results highlight the diverse spatial extents of dusty star-forming activity in galaxies, which are not solely concentrated into few dominant cores or clumps but are also distributed across extended disks.
These trends are broadly consistent with previous works based on high-resolution dust continuum maps on sub-kpc scales (\citealt{2016ApJ...829L..10I}; \citealt{2016ApJ...833..103H}; \citeyear{2019ApJ...876..130H};  \citealt{2018ApJ...859...12G}). It is reported that maps at $\gtrsim0.15^{\prime\prime}$ tend to recover most of the flux (\citealt{2016ApJ...833..103H}; \citealt{2019ApJ...876..130H}), while maps at $\lesssim0.05^{\prime\prime}$ often miss significant fractions of flux (\citealt{2016ApJ...829L..10I}; \citealt{2018ApJ...859...12G}). We also note that there are some caveats in investigating flux recovery. The first caveat is that 
the recovery rate should highly depend on sensitivity.
Another caveat is the use of the curve growth method. This method often results in ``over-recovering'' (e.g., \citealt{2018ApJ...859...12G}), possibly due to the contribution from the wing components of the synthesized beam and noise fluctuation, indicating that this effect may be prevalent (also for sources not showcasing over-recovering).

\subsection{ALMA Profile Fit} \label{sec:almafit}

\begin{deluxetable*}{lcccccccc}
\tabletypesize{\scriptsize}
\tablewidth{0pt} 
\tablecaption{ALMA 870\,$\mu$m profile measurements} \label{tab:ALMA_profile}
\tablehead{
\colhead{} & \multicolumn{4}{c}{Masked fits} & \multicolumn{4}{c}{Unmasked fits} \\
\hline
\colhead{ID} & \colhead{$n$} & \colhead{Re} & \colhead{b/a} & \colhead{P.A.} & \colhead{$n$} & \colhead{Re} & \colhead{b/a} & \colhead{P.A.} \\
\colhead{} & \colhead{} & \colhead{[kpc]} & \colhead{} & \colhead{[deg]} & \colhead{} & \colhead{[kpc]} & \colhead{} & \colhead{[deg]} 
}
\startdata 
ADF22.A1 & 0.94 $\pm$ 0.11 & 4.73 $\pm$ 0.53 & 0.42 $\pm$ 0.04 & 87.6 $\pm$ 8.8 & 3.78 $\pm$ 0.38 & 7.10 $\pm$ 1.07 & 0.33 $\pm$ 0.03 & 85.9 $\pm$ 8.6 \\
ADF22.A3 & 1.0 $\pm$ 0.16 & 2.29 $\pm$ 0.23 & 0.36 $\pm$ 0.04 & 76.9 $\pm$ 7.8 & 0.67 $\pm$ 0.14 & 2.06 $\pm$ 0.23 & 0.16 $\pm$ 0.02 & 75.5 $\pm$ 7.6 \\
ADF22.A4 & 1.27 $\pm$ 0.23 & 1.07 $\pm$ 0.15 & 1.0 $\pm$ 0.1 & 114.1 $\pm$ 100.5 & 2.04 $\pm$ 0.28 & 0.99 $\pm$ 0.15 & 0.86 $\pm$ 0.1 & 62.5 $\pm$ 12.6 \\
ADF22.A5 & 0.71 $\pm$ 0.1 & 2.59 $\pm$ 0.31 & 0.36 $\pm$ 0.04 & 74.3 $\pm$ 7.5 & 1.43 $\pm$ 0.16 & 2.59 $\pm$ 0.31 & 0.2 $\pm$ 0.02 & 73.1 $\pm$ 7.3 \\
ADF22.A6 & 0.73 $\pm$ 0.12 & 1.30 $\pm$ 0.15 & 0.72 $\pm$ 0.08 & 10.9 $\pm$ 4.0 & 1.42 $\pm$ 0.17 & 1.22 $\pm$ 0.15 & 0.63 $\pm$ 0.07 & 14.8 $\pm$ 4.3 \\
ADF22.A7 & 1.01 $\pm$ 0.17 & 1.53 $\pm$ 0.15 & 0.58 $\pm$ 0.06 & 17.4 $\pm$ 2.5 & 1.7 $\pm$ 0.22 & 1.45 $\pm$ 0.15 & 0.4 $\pm$ 0.04 & 18.4 $\pm$ 2.4 \\
\enddata
\end{deluxetable*}

\begin{figure*}
\epsscale{1.16}
\plotone{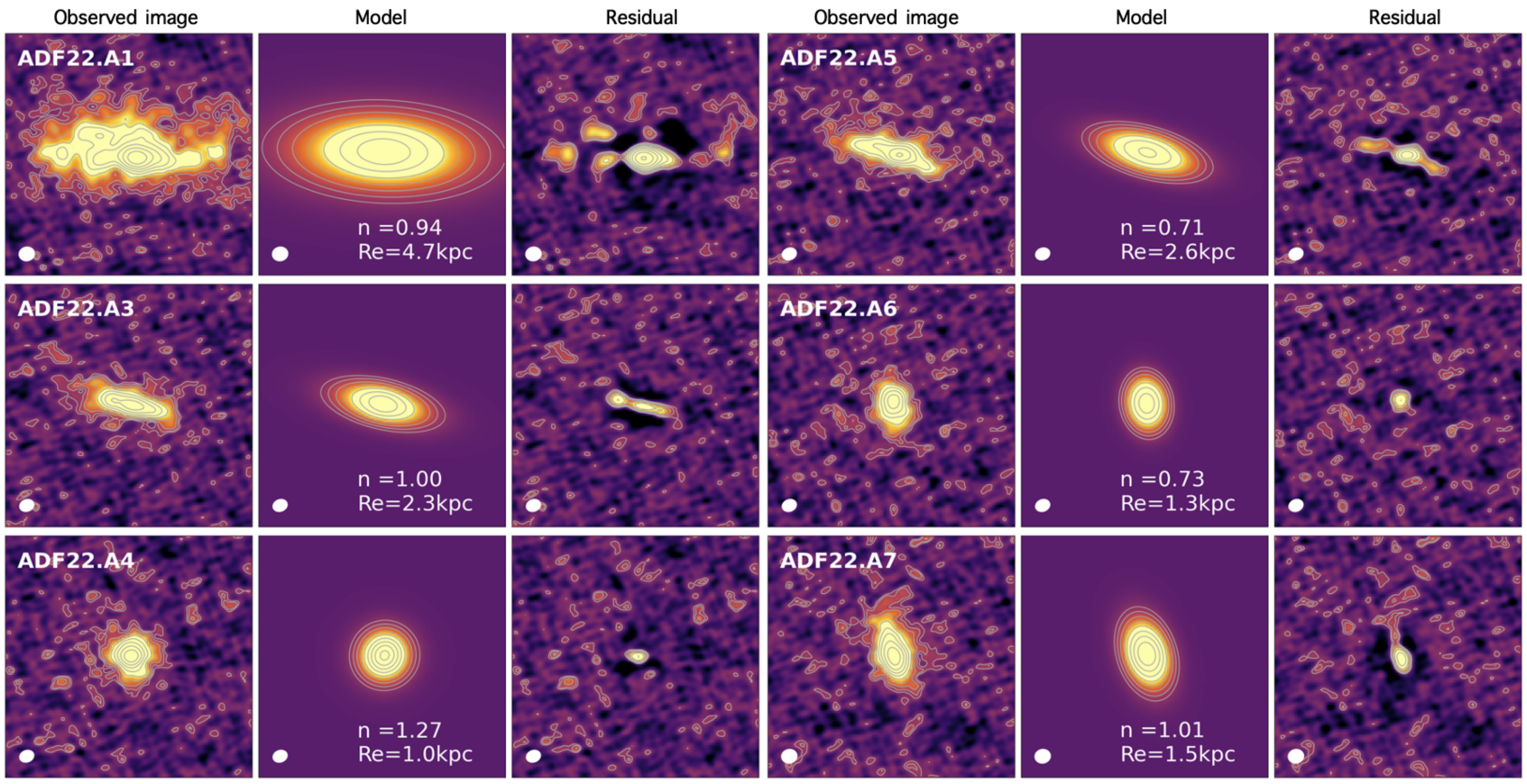}
\caption{
Results of the profile fit with ALMA 870\,$\mu$m imaging on the six brightest DSFGs in ADF22 (0.15$^{\prime\prime}$ resolution).
From left to right, each panel shows the observed images, the best-fit S\'ersic profile, and the residual maps. Each panel is $2.4^{\prime\prime}\times2.4^{\prime\prime}$ ($18.3\times18.3$\,kpc$^2$) in size.
Contours are $\sigma \times 1.5^n$, with $n=1$ to $n=10$.
We obtained the best-fit models (and residuals), masking residual pixels $>5\sigma$ iteratively as described in \S\ref{sec:almafit}.
Central cores and other substructures, suggestive of bars and clumps, are unveiled, together with the existence of disky profiles on a galaxy scale.
}
\label{fig:almafit}
\end{figure*}

The ALMA 870\,$\mu$m images taken at 0.15$^{\prime\prime}$ have sufficient spatial resolution to identify internal structures on spatial scales comparable to those observed with JWST/NIRCam. They also preserve the sensitivity to detect relatively extended emission thanks to the inclusion of data from the compact array, as the flux recovery rate suggests.
Hence we use these images to fit the surface brightness distribution of dust continuum emission in the DSFGs to quantitatively evaluate the profile of the dust continuum emission in these galaxies.
We adopt a S\'ersic profile (\citealt{1948AnAp...11..247D}; \citealt{1968adga.book.....S}), which is expressed as

\begin{equation}
    I(r) = I_e \exp \left( -\kappa \left[ \left( \frac{r}{r_e} \right)^{1/n} - 1 \right] \right),
\end{equation}

where $I(r)$ is the surface brightness at radius $r$, $I_e$ is the surface brightness at the effective radius $r_e$, $n$ is the S\'ersic index that determines the concentration of the profile, and $\kappa$ normalizes the profile such that the effective radius $r_e$ corresponds to the half-light radius, enclosing half of the total flux of the source within $r_e$.

First we fit a two-dimensional S\'ersic surface brightness profile, allowing the effective radius $r_e$, source position, axial ratio, position angle, and S\'ersic index $n$ to vary. To account for the beam convolution effect, the model is convolved with the synthesized beam, characterized by its semimajor axis, semiminor axis, and position angle as provided by the beam parameters of each map.
Following the recipe presented by \cite{2019ApJ...876..130H}, we performed the fit by iteratively masking residual pixels above $5\sigma$ until the masks converged. This step is suitable to appropriately characterize disk components in the presence of clumpy structures. This technique also ensures that any real positive structure in the disks does not artificially boost the fits of the underlying smooth profiles, which could otherwise perturb the results causing large negative troughs in the residual images (\citealt{2019ApJ...876..130H}). We also measured these parameters without masks to parameterize the whole profile including a concentrated core. 

The best-fit models and residual maps for the six brightest DSFGs in ADF22 are shown in Fig.~\ref{fig:almafit}. 
The derived parameters are summarized in Table~\ref{tab:ALMA_profile}, while the representative values of S\'ersic index and effective radius are also noted in the panels in Fig.~\ref{fig:almafit}. 
The derived S\'ersic index for the six bright DSFGs is $\langle n \rangle=0.86\pm0.05$ when we applied the bright emission masks. This suggests that an exponential disk profile is a common feature of the DSFGs. \cite{2016ApJ...833..103H} reported similar values for DSFGs in a general field of similar redshifts using comparable resolution images ($\langle n \rangle=0.9\pm0.2$), which indicates that such disky profiles of dust continuum emission are common characteristics among DSFGs in various environments.
The S\'ersic index increases when bright emission masks are not used ($\langle n \rangle=1.37\pm0.08$). This suggests that the DSFGs have an extra component concentrated into a core, which is also recognizable in the residual maps and detections in the 1.1\,mm maps with higher angular resolutions. An exception is ADF22.A3, which shows $n=0.67\pm0.14$ without masks (while it has $n=1.00\pm0.16$ with masks). One possible explanation is that a bar, which tends to have a flatter profile, dominates the bright part. 

The effective radius for the six DSFGs ranges from $0.14^{\prime\prime}$ to $0.62^{\prime\prime}$ (1.06 to 4.73\,kpc), showcasing a variety of the disk sizes as reported by previous works for general field DSFGs (e.g., \citealt{2019MNRAS.490.4956G}). There is also an exception. ADF22.A1 has a remarkably large dust disk ($r=0.62^{\prime\prime}$ , 4.73\,kpc) also associated with large gas and stellar disks (\citealt{2024arXiv241022155U}), which is larger than any of previously known samples observed at a comparable resolution (\citealt{2019MNRAS.490.4956G}).
The special environment, associated with gas filaments in the proto-cluster core, can account for such a huge disk (\citealt{2024arXiv241022155U}).
A remarkable finding is the prevalence of substructures in the bright DSFGs. This could reflect bulges, bars, and clumps in spiral arms, as also reported by \cite{2019ApJ...876..130H} for DSFGs in general fields. We will discuss the origin more in \S \ref{sec:discussion_origin} together with NIRCam data and ALMA Band-6 images at 0.05$^{\prime\prime}$--0.08$^{\prime\prime}$. 

As a test, we refit the sample with a fixed $n$ ($n=1$) without masking. The results are shown in Fig.~\ref{fig:almafit_n1}. 
For the six bright DSFGs, this fit broadly yielded similar parameters compared with the fit with bright emission masks. This may be reasonable since the derived S\'ersic index with masks are close to $n=1$. In the following, we take the fits with masks as our primary estimates. 
For two faint sources detected near the primary targets, ADF22.10 and ADF22.11, fit with a S\'ersic index of $n = 1.0$ work well (see Appendix), while this assumption is necessary due to the low significance of the detected emission. The final source ADF22.A16 was excluded from all fits since it is too faint at this resolution (Fig.~\ref{fig:almamaps_faint}).

We note that high fidelity data is required for high resolution profile fits which also need to capture extended emission from disks. This is showcased in our data set. The ALMA Band-6 data is sensitive to compact emission but does not have sufficient sensitivity for disk profile characterizations. In these data one-component fits without masks sometimes resulted in high $n$ values ($n\gtrsim5$). This does not mean that the galaxy does not have dust continuum emission from a disk (this is caused by the lacking of the sensitivity for extended components), as also suggested by stacking analysis by \cite{2019MNRAS.490.4956G}. However, such high-resolution images offer a finer views of the galactic structures, which is useful to resolve small scale structures.

\section{Stellar morphologies and comparison with dust continuum} \label{sec:stellar_mor}

\subsection{Stellar and dust emission in DSFGs}

\begin{figure*}
\epsscale{1.16}
\plotone{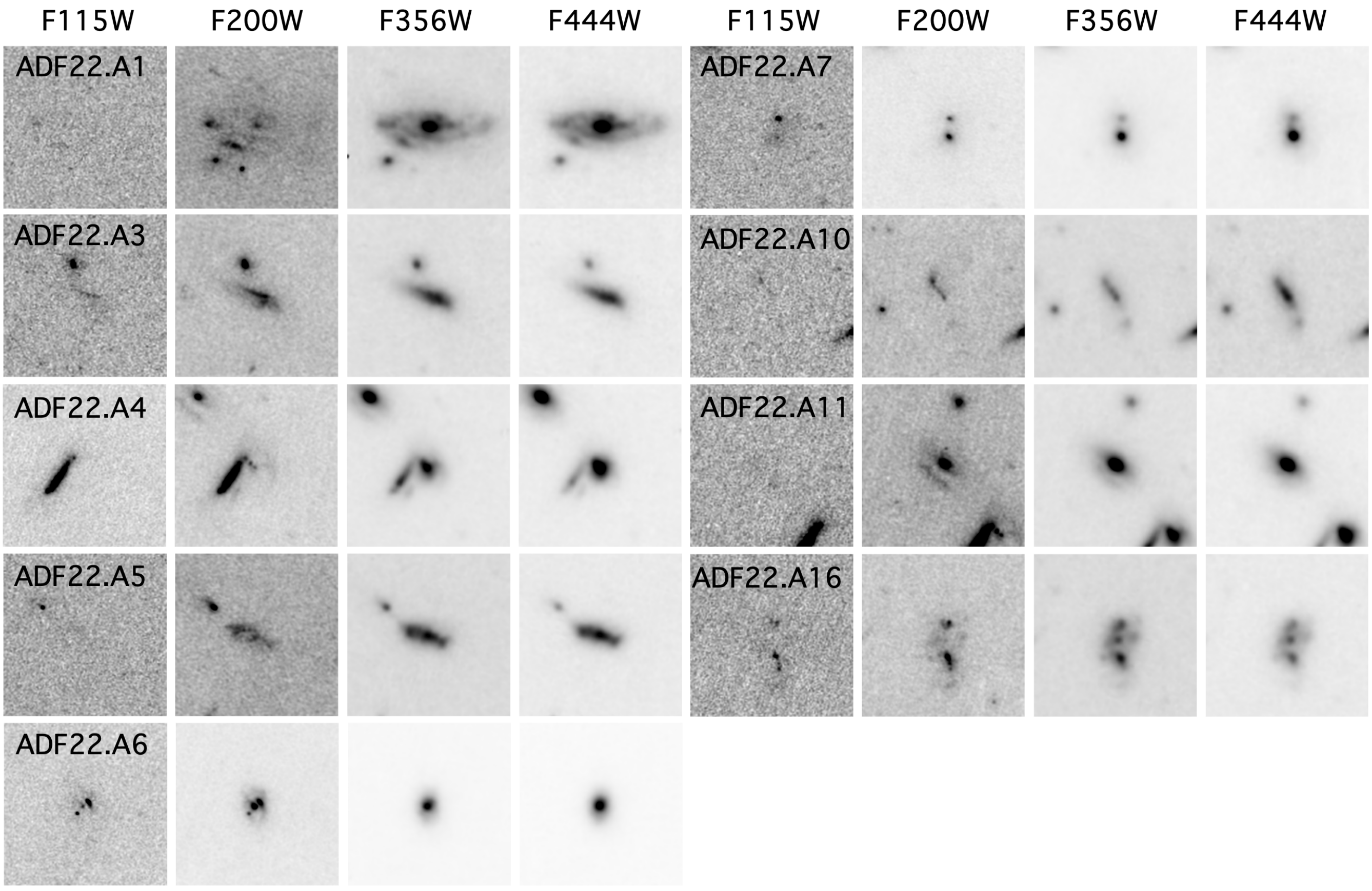}
\caption{
Postage stamps of the NIRCam images as marked. Each panel is $3.8^{\prime\prime}\times3.8^{\prime\prime}$ ($29\times29$\,kpc$^2$ at $z=3.09$) in size. Dust extinction is severe in bluer bands (F115W and F200W).
}
\label{fig:nircam}
\end{figure*}

\begin{figure*}
\epsscale{1.16}
\plotone{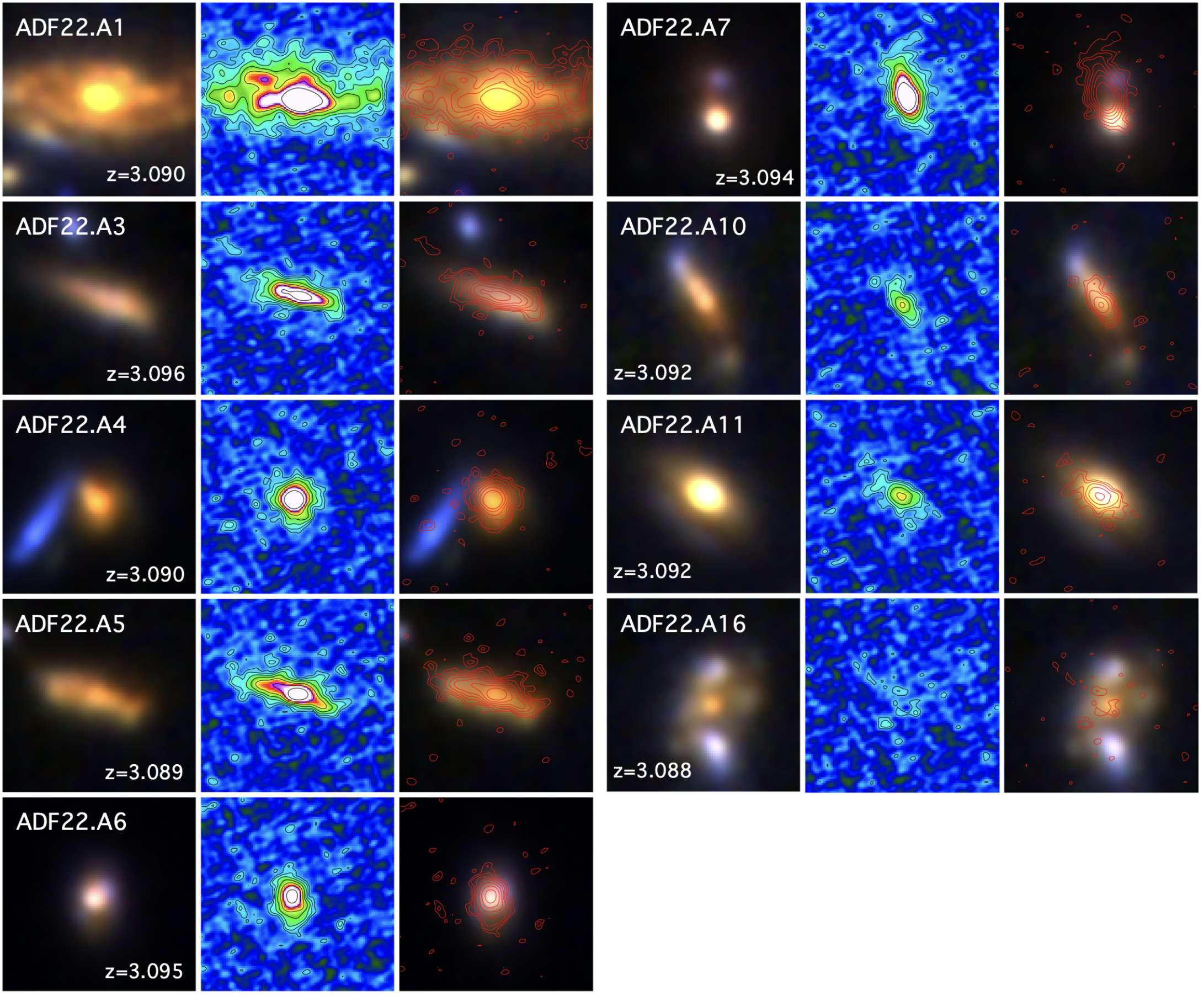}
\caption{
A combined ALMA and JWST views of nine DSFGs. (left) NIRCam pseudo-color images (blue: F200W, green: F356W, red: F444W) for the nine proto-cluster DSFGs in ADF22, in which PSF sizes are matched to that of F444W. Each panel is $2^{\prime\prime} \times 2^{\prime\prime}$ in size. 
(middle) ALMA 870\,$\mu$m images. Contours represent $\sigma \times 1.5^n$ ($n$=2 to 11) of the $870\mu$m emission.
(right) NIRCam images with the ALMA contours.
JWST/NIRCam reveals diverse stellar colors and morphologies among the DSFGs.
The dust continuum emission observed with ALMA is broadly coincident with the stellar light.
}
\label{fig:nircam_alma}
\end{figure*}

The JWST/NIRCam images provide the opportunity to resolve the internal structures of DSFGs and associate them with dust continuum emission. Fig.~\ref{fig:nircam} shows the images in the individual NIRCam filters. A notable trend is that a significant fraction of DSFGs are heavily obscured, rendering most regions invisiblein the bluer bands even in the $K_s$-band (F200W). This trend highlights that a wavelength of approximately 5000\,\AA~in the rest frame is too short to effectively trace stellar emission even from massive galaxies in such a dusty population at $z\sim$. While this wavelength is often considered suitable for deriving stellar mass distributions due to its inclusion in the rest-frame optical regime, its limitations in dusty systems must be acknowledged.

In Fig.~\ref{fig:nircam_alma}, we presents a side-by-side comparison of NIRCam color images, which combine data from the F200W, F356W, and F444W filters, and 870\,$\mu$m maps for proto-cluster DSFGs. This reveals a striking diversity in the size, color, inclination, and structure of the DSFGs.
In most cases (eight out of nine DSFGs), the peaks of 870\,$\mu$m emission broadly coincide with those of the rest-frame $\sim1.1$\,$\mu$m emission in the F444W images at the $\sim0.15^{\prime\prime}$ resolution. This indicates that we are beginning to directly capture stellar mass assembly in the central regions of galaxies, where dust extinction is most severe (\citealt{2024A&A...691A.299G}; \citealt{2025ApJ...978..165H}). The exception is ADF22.A7, which shows an offset between the peaks of 870\,$\mu$m and F444W emissions. This offset is intrinsic and not due to astrometric uncertainties (both A1 and A7 are in the same single FoV of ALMA Band7/Band6 observations, and astrometry is calibrated across the ADF22 field). We will discuss the nature of this DSFG in \S \ref{sec:discussion_environment}.

The bluer band images (F115W and F200W, $\sim$2800\AA~and $\sim$4900\AA~in the restframe) also provide several new insights. Some DSFGs, such as ADF22.A1 and ADF22.A10, exhibit blue arms in their outskirts, suggesting the presence of less dusty regions within these DSFGs as well as inhomogeneous dust obscuration across their disks.
Several clumps are visible in the shorter wavelength images, as exemplified by ADF22.A4 and ADF22.A6 (Fig.~\ref{fig:nircam}, see also Fig.~\ref{fig:f200w_870} in Appendix). These features are not detectable in the F356W/F444W maps, likely due to their intrinsically blue(and lower masses) nature and/or the lower resolution of the longer wavelength images.
The blue clumps that spatially overlap with the 870\,$\mu$m (and F444W) emission possibly reflect nuclear activity, including nuclear starbursts and AGNs (both DSFGs host X-ray-detected AGNs, while their SEDs are dominated by galaxies (\cite{2023ApJ...951...15M}). The remaining clumps may represent young star clusters, similar to the UV clumps observed by HST in $z=1-2$ galaxies (e.g., \citealt{2004ApJ...600L.139C}; \citealt{2005ApJ...627..632E}; \citealt{2007ApJ...658..763E}; \citeyear{2009ApJ...692...12E}; \citealt{2008A&A...486..741B}; \citealt{2008ApJ...687...59G}; \citeyear{2011ApJ...733..101G}). 

\subsection{NIRCam Profile Fit}

\begin{deluxetable*}{lcccccccc}
\tabletypesize{\scriptsize}
\tablewidth{0pt} 
\tablecaption{NIRCam F444W profile measurements} \label{tab:f444w_profile}
\tablehead{
\colhead{} & \multicolumn{4}{c}{Free $n$ fits} & \multicolumn{4}{c}{Fixed $n=1$ fits} \\
\hline
\colhead{ID} & \colhead{$n$} & \colhead{Re} & \colhead{b/a} & \colhead{P.A.} & \colhead{$n$} & \colhead{Re} & \colhead{b/a} & \colhead{P.A.} \\
\colhead{} & \colhead{} & \colhead{[kpc]} & \colhead{} & \colhead{[deg]} & \colhead{} & \colhead{[kpc]} & \colhead{} & \colhead{[deg]} 
}
\startdata 
ADF22.A1 & 1.61 $\pm$ 0.17 & 6.68 $\pm$ 0.69 & 0.39 $\pm$ 0.04 & 89.8 $\pm$ 8.9 & 1.00 & 5.86 $\pm$ 0.60 & 0.39 $\pm$ 0.04 & 89.8 $\pm$ 8.9 \\
ADF22.A3 & 0.71 $\pm$ 0.09 & 3.49 $\pm$ 0.37 & 0.31 $\pm$ 0.03 & 68.9 $\pm$ 6.9 & 1.00 & 3.77 $\pm$ 0.39 & 0.30 $\pm$ 0.03 & 68.7 $\pm$ 6.9 \\
ADF22.A4 & 1.59 $\pm$ 0.29 & 1.81 $\pm$ 0.19 & 0.86 $\pm$ 0.11 & 48.2 $\pm$ 9.2 & 1.00 & 1.60 $\pm$ 0.16 & 0.86 $\pm$ 0.09 & 46.7 $\pm$ 5.2 \\
ADF22.A5 & 0.36 $\pm$ 0.04 & 3.28 $\pm$ 0.34 & 0.31 $\pm$ 0.04 & 73.5 $\pm$ 7.3 & 1.00 & 3.76 $\pm$ 0.39 & 0.30 $\pm$ 0.03 & 73.4 $\pm$ 7.3 \\
ADF22.A6 & 1.43 $\pm$ 0.38 & 1.50 $\pm$ 0.15 & 0.69 $\pm$ 0.13 & -4.9 $\pm$ 17.5 & 1.00 & 1.41 $\pm$ 0.14 & 0.70 $\pm$ 0.11 & -5.2 $\pm$ 11.2 \\
ADF22.A7 & 3.21 $\pm$ 0.37 & 2.24 $\pm$ 0.25 & 0.54 $\pm$ 0.06 & 10.6 $\pm$ 1.2 & 1.00 & 1.55 $\pm$ 0.17 & 0.57 $\pm$ 0.07 & 10.5 $\pm$ 1.2 \\
ADF22.A10 & 1.14 $\pm$ 0.14 & 3.33 $\pm$ 0.38 & 0.28 $\pm$ 0.06 & 27.8 $\pm$ 4.9 & 1.00 & 3.22 $\pm$ 0.37 & 0.28 $\pm$ 0.05 & 27.9 $\pm$ 3.7 \\
ADF22.A11 & 3.51 $\pm$ 0.46 & 2.83 $\pm$ 0.30 & 0.54 $\pm$ 0.08 & 54.1 $\pm$ 12.1 & 1.00 & 1.62 $\pm$ 0.17 & 0.57 $\pm$ 0.07 & 55.0 $\pm$ 5.9 \\
ADF22.A16 & 0.42 $\pm$ 0.11 & 3.80 $\pm$ 0.50 & 0.44 $\pm$ 0.07 & -3.3 $\pm$ 5.8 & 1.00 & 4.17 $\pm$ 0.60 & 0.41 $\pm$ 0.05 & -3.4 $\pm$ 6.6 \\
\enddata
\end{deluxetable*}

\begin{figure*}
\epsscale{1.0}
\plotone{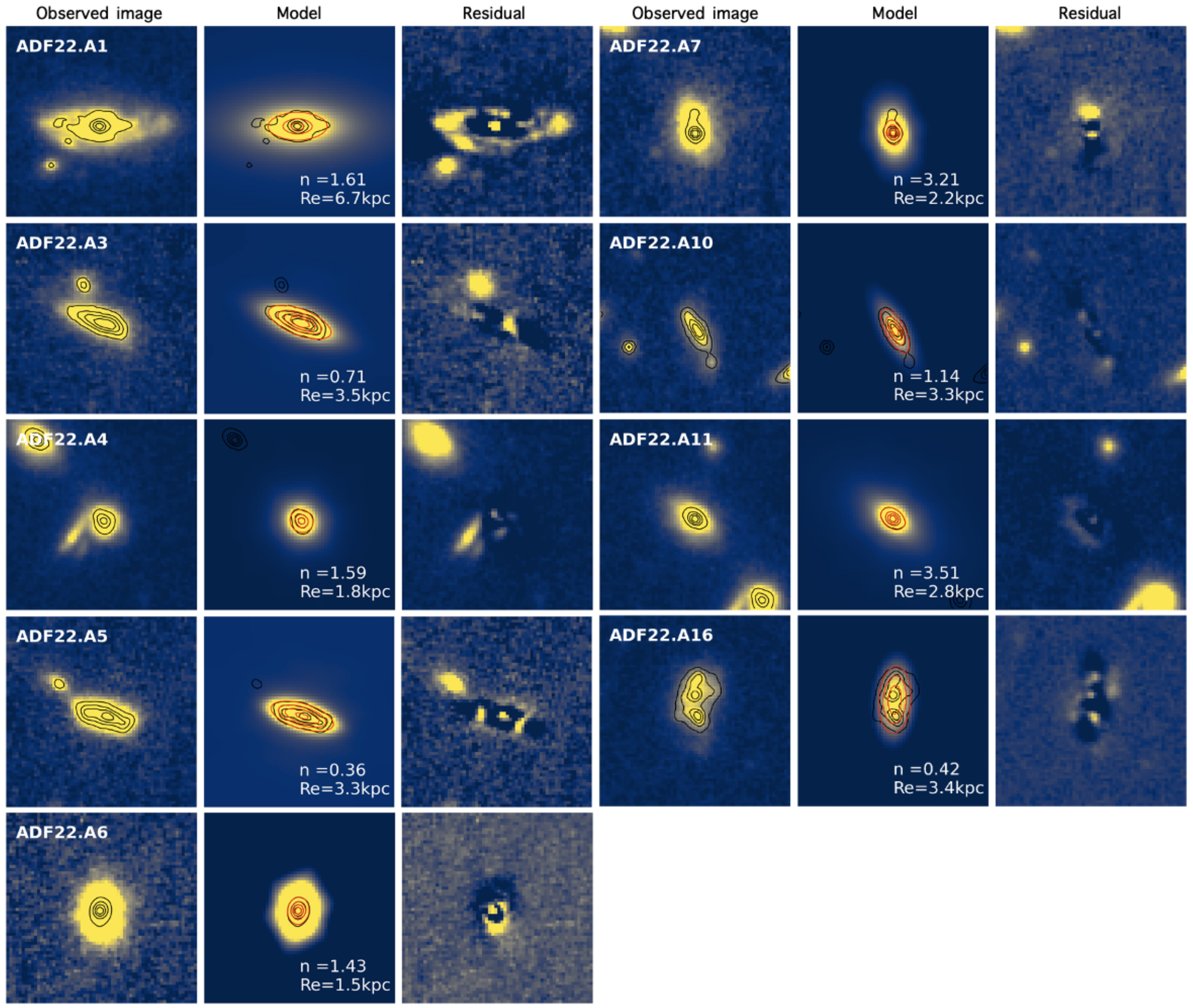}
\caption{
Postage stamps of JWST NIRCam F444W images for the nine DSFGs in ADF22 ($30\times30$\,kpc$^2$ in size). As in Fig.~\ref{fig:almafit}, each panel from left to right displays the observed image, the best-fit S\'ersic profile, and the residual map. Black (red) contours represent 20\%, 40\%, 60\%, and 80\% of the peak flux of the observed (modeled) images. The majority of the profiles tend to show $n>1$, suggesting the presence of a forming bulge. 
}
\label{fig:f444wfit}
\end{figure*}

The NIRCam F444W images were utilized to characterize morphologies in stellar emission since they provide the longest wavelength coverage among the NIRCam data. The rest-frame wavelength at $z = 3.09$ is approximately 1.1\,$\mu$m, corresponding to the rest-frame near-infrared. Thus, the F444W image is optimal for tracing stellar structures among the available images, while we note that the rest-frame wavelength may still be affected by dust extinction (e.g., \citealt{2025ApJ...978..165H}; \citealt{2024arXiv241022155U}). 

We fit a S\'ersic profile to the F444W surface brightness of individual DSFGs using {\sc galfit} \citep{2002AJ....124..266P, 2010AJ....139.2097P}. The input files for the {\sc galfit} analysis include a science image, a PSF map, a sigma map, and a mask map. For the F444W image, we use the original pixel scale (0.063$^{\prime\prime}$\,pixel$^{-1}$) without applying drizzling, and we create 150\,pixel\,$\times$\,150\,pixel thumbnails centered on each DSFG.
We obtain the most realistic PSFs following previous studies (\citealt{2022ApJ...939L...7C}; \citealt{2025ApJ...978..165H}). 
Using WebbPSF (\citealt{2014SPIE.9143E..3XP}), we generated a PSF matched to in-flight JWST data. We use the PSF simulated for our instrument setup and date of observation.
The {\sf charge diffusion sigma} option of the WebbPSF is adjusted to fit the PSF to nearby unsaturated stars.
The input image values are converted to have units of count rate and the sigma image is generated internally by {\sc galfit}.
The mask image is generated using {\sc sextractor} (\citealt{1996A&AS..117..393B}). Sources detected with $>1.5$ sigma for $>12$ adjacent pixels are extracted and masked. Measurements are iterated 1000 times with a range of initial values, which finally provides the representative values and errors as median and standard deviation. 
We estimate the errors for each output parameter by adding a 10\% fractional uncertainty in quadrature to all parameters, following the approach of \citet{2025ApJ...978..165H}.
Fig.~\ref{fig:f444wfit} shows the best-fit model and residual images, together with the observed F444W images, for the six bright ADF22 DSFGs. Similar to 870 $\mu$m image, fits with {\sc galfit} are also performed with a fixed $n=1$.
The derived parameters are summarized in Table~\ref{tab:f444w_profile}.

\subsection{Comparison between ALMA and JWST profiles}

\begin{figure}
\epsscale{1.16}
\plotone{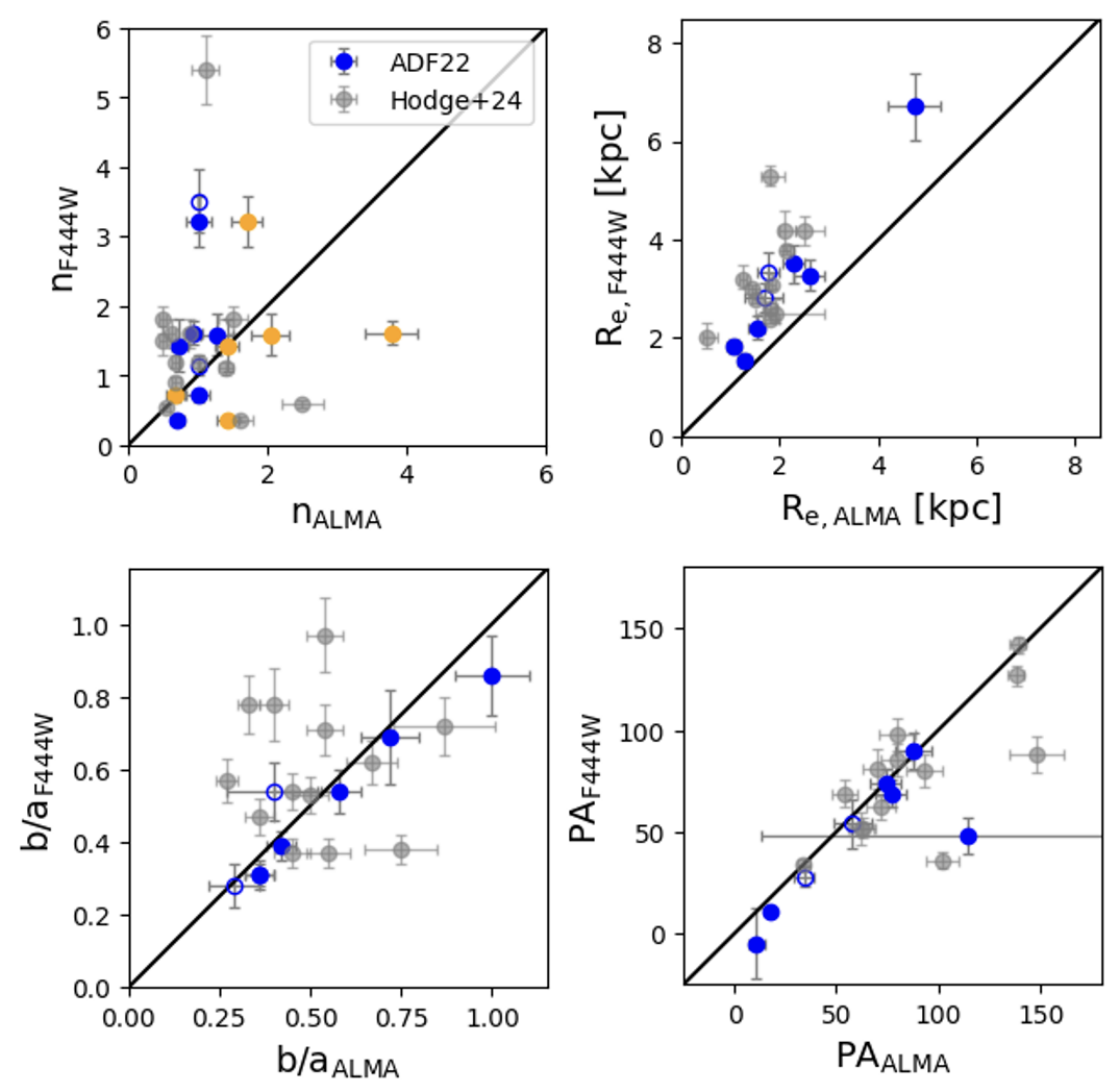}
\caption{
Comparison of parameters derived by profile fit on ALMA 870\,$\mu$m and JWST NIRCam F444W images. Blue filled circles show the results from ALMA profile fit with the bright source masks, while yellow circles show the results without the masks (only in the top right panel). Open blue circles shows ADF22.A10 and A11, whose measurements were obtained with $n=1$ and adopting no masks. For comparison, measurements from previous works (\citealt{2025ApJ...978..165H}) also plotted.
The axis ratios and position angles measured by the two ways show good consistency, supporting the scenario that dust emission comes from across disks.
}
\label{fig:profile_comp}
\end{figure}

The derived parameters characterizing the dust and stellar emission profiles are compared in Fig.~\ref{fig:profile_comp}. For reference, we also include measurements from $z\sim3$ DSFGs originally identified in the general field from \cite{2025ApJ...978..165H}.
The S\'ersic index, $n$, show moderate correlation within the range of $n\sim1-2$ in most cases. The galaxies with $n_{\rm F444W}>1$ can be interpreted as a mixture of disks and central spheroidal components. There are two sources with $n_{\rm F444W}<1$ (ADF22.A3 and ADF22.A5), which may be attributed by flattened profile due to significant dust extinction and/or possibly bars. If bright emission masks were not applied, $n_{\rm ALMA}$ would exhibit larger values, reflecting the influence of bright and relatively compact cores in dust emission of some sources. A similar situation likely applies to F444W, where we conservatively adopted a single-component fit due to the limited sensitivity. There is an exception, ADF22.A7, which shows
a remarkable discrepancy between the two types of measurements. We discuss this source more in \S \ref{sec:discussion_environment}. 

For the effective radius, $R_{\rm e}$, the F444W measurements tend to be larger than those of dust, although a positive correlation exists between the two. This trend is broadly consistent with previous studies (e.g., \citealt{2022ApJ...939L...7C}; \citealt{2025ApJ...978..165H}; \citealt{2024A&A...691A.299G}). Notably, the trend remains unchanged when adopting the effective radius measured with a fixed S\'ersic index ($n=1$). From the perspective of hydrodynamical simulations, \cite{2022MNRAS.510.3321P} predicted smaller stellar mass sizes compared to 850\,$\mu$m sizes in massive star-forming galaxies at $z=1-5$. 
On the other hand, observational studies have reported the opposite behavior in observed sizes measured at submillimeter wavelengths and F444W (e.g., \citealt{2022ApJ...939L...7C}; \citealt{2024A&A...691A.299G}). Several factors could contribute to this apparent contradiction. For instance, simulations may require improvements in accounting for details of central mass growth and obscuration associated with bulge formation. It should be also noted that F444W measurements can still be affected by dust extinction, which may flatten the observed profiles (decreasing the measured $n$). As a result, the intrinsic stellar mass profile could be more compact than it appears in F444W images.

We can see good consistency between the ALMA and F444W measurements on the b/a ratio and position angle (PA). This result is inline with the idea that dust emission generally originates from across disks.
To evaluate the relationship between the parameters derived from the F444W and 870\,$\mu$m data, we compute the Pearson correlation coefficient ($\rho$) following \cite{2025ApJ...978..165H},  which quantifies the strength of a linear monotonic association between two datasets, for the six brightest DSFGs. The resulting values of $\rho$ (along with their corresponding probabilities) are 0.37 (0.47) for $n$, 0.99 (0.00) for $R_e$, 0.99 (0.00) for b/a, and +0.79 (0.06) for PA. The results again indicate the significant correlation except for the S\'ersic index.

The results and trends shown in Fig.~\ref{fig:profile_comp} are broadly consistent with those for DSFGs in general fields reported by \cite{2025ApJ...978..165H} (see also \citealt{2024A&A...691A.299G}). However, a notable exception arises in the case of the axis ratio. While they found no correlation between the ALMA and F444W measurements for the axis ratio, the ADF22 DSFGs exhibit a clear correlation. 
Since all of the DSFGs presented in this paper are located in the proto-cluster core, this might be linked to the formation process preferentially occurred in the biased environment (we will revisit this point in the next section).
Furthermore, the comparison between the two studies highlights the unique large stellar and dust sizes of ADF22.A1 in the proto-cluster core (\citealt{2024arXiv241022155U}).

\section{Discussion}

\subsection{The origins of dust continuum emission} \label{sec:discussion_origin}

\begin{figure}
\epsscale{1.16}
\plotone{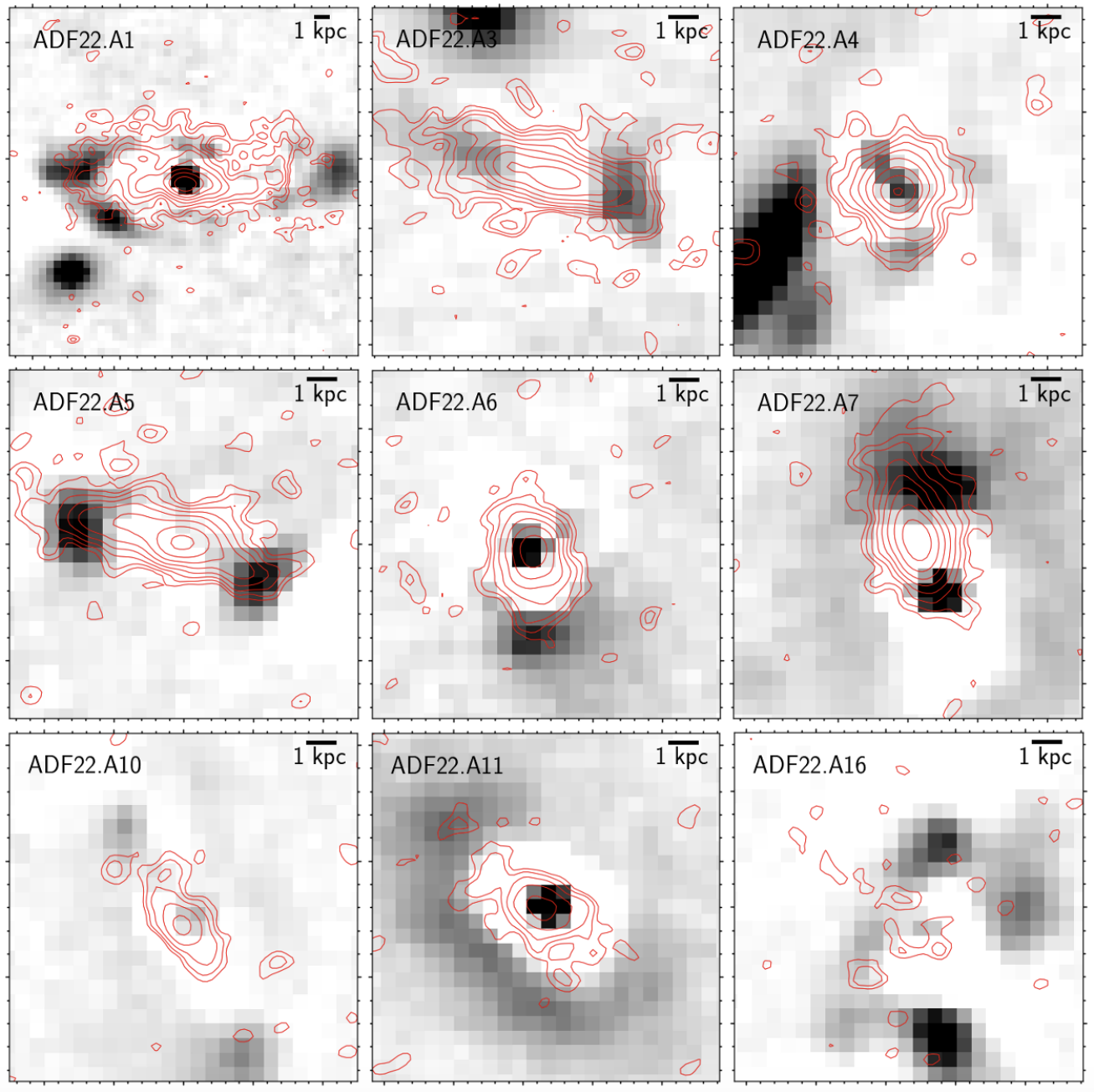}
\caption{
The F444W residual images after subtracting the best-fit S\'ersic model with a fixed S\'ersic index ($n=1$). The contours represent the 870\,$\mu$m emission at 0.15$^{\prime\prime}$, as shown in Fig.~\ref{fig:almamaps}. 
The peaks of the dust continuum often coincide with the central residuals of the F444W emission, suggesting the co-existence of stellar spheroids and central starbursts in the DSFGs.
}
\label{fig:f444residual}
\end{figure}

Multi-scale ALMA submm images, combined with rest-frame optical to near-infrared NIRCam images, provide critical insights into identifying the structures responsible for dust emission in galaxies on a resolved scale.
The emission, after masking the bright regions, is well characterized by a disk-like profile, with the axis ratios and position angles showing good consistency, as described above. These observational clues suggest that the dust continuum originates predominantly across disks in DSFGs. This interpretation is supported by the stacked detection of faint and extended components reported by \cite{2019MNRAS.490.4956G}.
The dust emission is also characterized by dominant bright cores, a feature reported in numerous previous studies (e.g., \citealt{2015ApJ...799...81S}; \citealt{2015ApJ...810..133I}; \citealt{2016ApJ...833..103H}; \citealt{2019MNRAS.490.4956G}; \citealt{2024Natur.636...69T}). The NIRCam images now reveal that these cores are situated at the centers of disk-like galaxies. 
Fig.~\ref{fig:f444residual} compares the the F444W residual images, which is obtained subtracting a $n=1$ best-fit model, with the 870\,$\mu$m emission at $0.15^{\prime\prime}$ for the nine ADF22 DSFGs at $z\approx3.09$. A notable feature is that about a half of the F444W residual image have residuals at the center, which corresponds the peak of the dust continuum emission (ADF22.A1, A4, A6, A11). This result suggests concurrent starburst activity and, to a certain extent, the presence of mature stellar bulges. Similar situation can be found in other DSFGs in a general field, such as ALESS3.1 at $z=3.375$ (\citealt{2025ApJ...978..165H}), and hence can be a general nature of DSFGs. This point will be explored further below.

Another notable feature is that the combination of disks and cores alone is insufficient to fully characterize the observed profiles. One key characteristic is the presence of dusty bars (\citealt{2019ApJ...876..130H}; \citealt{2024MNRAS.527.8941T}), most prominently seen in ADF22.A1 (\citealt{2024arXiv241022155U}). ADF22.A3 and ADF22.A5 also exhibit similar elongated structures in the high-resolution dust emission images at 0.08$^{\prime\prime}$ (Fig.~\ref{fig:almamaps}) and residual images (Fig.~\ref{fig:almafit}), which reveal structures resembling the offset ridges in local barred galaxies. These may reflect the prevalence of bars in DSFGs (e.g., \citealt{2023ApJ...958...36S}; \citealt{2024arXiv240401918A}), though kinematic data may be required to confirm this scenario, as demonstrated for ADF22.A1 (\citealt{2024arXiv241022155U}), and it is also required to expand the sample size.
There are no clear counterparts for these bars (or bar-like features) in the NIRCam images (e.g., Fig.~\ref{fig:f444residual}). The most significant cause would be dust obscuration, could easily obscure bar structures in DSFGs (\citealt{2017ApJ...839...58S}; \citealt{2025ApJ...978..165H}; \citealt{2024arXiv241022155U}).
An intriguing feature is the S\'ersic indices of the two DSFGs, ADF22.A3 and ADF22.A5. These  exhibit $n<1$ in the F444W profile fit. The flattened emission profiles may result from significant dust extinction, as discussed above, but could also be caused by the presence of bars. Bars can exhibit have lower S\'ersic indices in local galaxies (e.g., \citealt{2018MNRAS.473.4731K}). Both DSFGs exhibit an edge-on-like geometry, and if bars dominate the stellar emission, the measured S\'ersic indices could fall below unity. It is prevalent for inclined disky galaxies to exhibit pairs of F444W residuals near the edge of the 870\,$\mu$m emission (e.g., ADF22.A1, A3, A5, A10; see Fig.~\ref{fig:f444residual}). As demonstrated in the case of ADF22.A1, this can be associated with stellar spiral arms.

Additional dusty clumps (not in the core) are observed in ADF22.A1 (Fig.~\ref{fig:almamaps}, Fig.~\ref{fig:almafit}), suggesting that such clumps, alongside cores and bulges, exist in at least some DSFGs (\citealt{2016ApJ...829L..10I}; \citealt{2018Natur.560..613T}; \citealt{2019ApJ...876..130H}), while we need more data to derive their prevalence and nature.
In summary, the dust continuum emission observed in DSFGs cannot be described by simple structures, such as a single compact bright component, but is instead attributed to multiple substructures, including disks, cores, bars, and clumps.

\subsection{Bulge formation and morphological transformation in DSFGs} \label{sec:discussion_bulge}

\begin{figure}
\epsscale{1.16}
\plotone{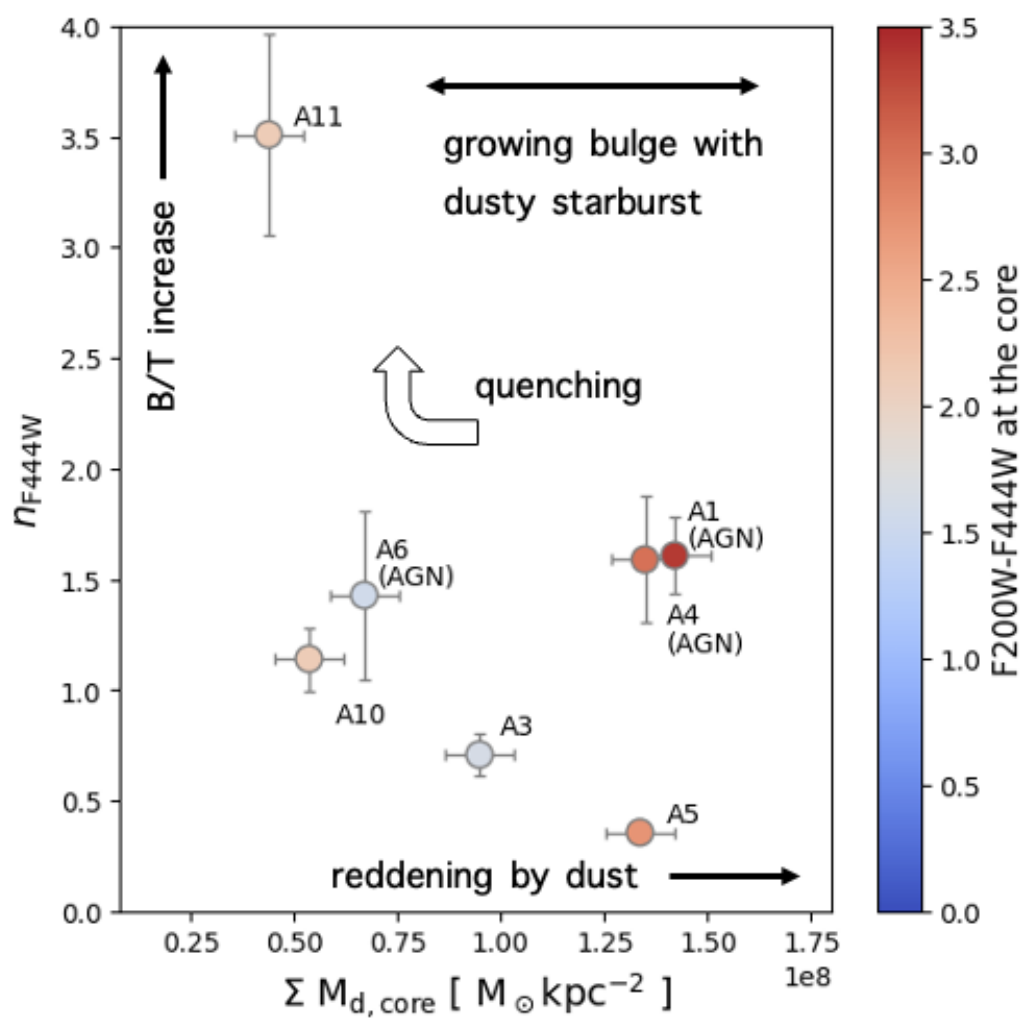}
\caption{
The relation among dust mass surface density at the core ($\Sigma$\,M$_{\rm d, core}$),  S\'ersic index measured for the F444W image ($n_{\rm F444W}$), and F200W-F444W color at the core. Dusty starburst at the core, which causes severe dust reddening, would correspond to the rapid growth phase of bulges.
}
\label{fig:md_n}
\end{figure}

Leveraging the unique advantage of our observations that the DSFGs are well-resolved in both stellar and dust continuum emission, we aim to bridge these two perspectives to shed light on the process of bulge formation.
One particularly exciting discovery is the presence of compact dusty cores in the bright DSFGs, which are strikingly prominent at a resolution of 350\,pc (Fig.~\ref{fig:almamaps}). To quantify this, we estimated the dust mass surface density at the core using the 0.05$^{\prime\prime}$ resolution 1.1\,mm images, as described below:

\begin{equation}
\Sigma M_{\mathrm{d, core}} = \frac{S_{\nu_{\mathrm{obs, 0.05^{\prime\prime}}}} D_L^2 (1+z)^{-(3+\beta)}}{\kappa(\nu_{\mathrm{ref}}) B_\nu(\nu_{\mathrm{ref}}, T_{\mathrm{d}}) A_{\mathrm {beam}}} \left(\frac{\nu_{\mathrm{ref}}}{\nu_{\mathrm{obs}}}\right)^{2+\beta}
\end{equation}

Here, $S_{\nu_{\mathrm{obs, 0.05^{\prime\prime}}}}$ is the peak flux density measured at 0.05$^{\prime\prime}$ resolution, and $D_L$ is the luminosity distance. The terms $\nu_{\mathrm{ref}}$ and $\nu_{\mathrm{obs}}$ represent the reference frequency and observed frequency, respectively.
We adopt a spectral emissivity index of $\beta = 1.8$, a dust mass absorption coefficient at 850\,$\mu$m of $\kappa(\nu_{\mathrm{850}}) = 0.04 \, \mathrm{m}^2 \, \mathrm{kg}^{-1}$ \citep{2001ApJ...554..778L}, and a mass-weighted dust temperature of $T_{\mathrm{d}} = 25 \,\mathrm{K}$ \citep{2016ApJ...820...83S}. The term $B_\nu(\nu_{\mathrm{ref}}, T_{\mathrm{dust}})$ denotes the Planck function, and $A_{\rm beam}$ is the beam solid angle.
The effect of the cosmic microwave background (CMB) is not corrected for, as it is not dominant at the studied redshift \citep{2013ApJ...766...13D}.

To evaluate the reddening effect, we also performed photometry at the core using an aperture centered on the dusty core with $d=0.15^{\prime\prime}$ and measured the color F200W$-$F444W in AB magnitude. Together with the S\'ersic index measured for the F444W image ($n_{\rm F444W}$), we plot the relation among the three parameters: dust mass, stellar morphology, and reddening in rest-frame optical-to-near-infrared color, in Fig.~\ref{fig:md_n}. ADF22.A7 is excluded since the configuration of dust and stellar emission is complicated in this DSFG.

One trend we can see in the diagram is the correlation between dust mass surface density and color. The three DSFGs, ADF22.A1, A4, and A5, hosting the brightest dusty cores, show the reddest color, highlighting that dust obscuration is a primary cause of the reddening of the SED in this wavelength range (\citealt{2024A&A...691A.299G}; \citealt{2025ApJ...978..165H}).
Two of them (ADF22.A1 and A4) also show relatively high S\'ersic indices, $n_{\rm F444W} \approx 1.5 - 2$. Together with the spatial overlap of core emissions between the dust and stellar components (Fig.~\ref{fig:nircam_alma}), this indicates that the DSFGs may have already assembled significant stellar mass in the proto-bulge, while the phase of bursting growth is still ongoing. The low index, $n_{\rm F444W} \lesssim 0.5$, observed in ADF22.A5 can be attributed to a combination of (i) severe dust extinction, (ii) relatively immature stellar mass in the proto-bulge, and (iii) a possible bar structure. Although it is challenging to distinguish these scenarios at the moment, the growth of proto-bulges may increase $n$ values during the DSFG phase.
The result also suggests that the near-infrared-dark nature observed in some DSFGs (e.g., \citealt{2014ApJ...788..125S}; \citealt{2020A&A...640L...8U}; \citealt{2022A&A...659A.154I}) is associated with the rapid growth phase of bulges (\citealt{2021MNRAS.502.3426S}).

Interestingly, ADF22.A1 exhibits a bar structure identified in both dust and [C\,{\sc ii}] emission (\citealt{2024arXiv241022155U}), and ADF22.A5 also shows signatures of offset ridges in the dust continuum (Fig.~\ref{fig:almamaps}). These clues suggest that bars play a key role in driving gas to the central regions, fueling mass growth in the core and triggering dusty starburst activity.
It is also found that the brightest, compact dusty cores are not ubiquitous among DSFGs. In the ADF22 sample, ADF22.A3 and A6 exhibit lower dust mass surface densities and consequently show slightly more moderate reddening in their colors. This diversity might reflect different stages along a similar evolutionary path (i.e., they are relatively immature and will develop bursty cores in the future, or they previously hosted the brightest cores but have now transitioned out of the rapid growth phase), or it could indicate multiple channels for DSFG formation (e.g., bypassing the bursty phase entirely). Statistical samples and detailed analyses of SEDs will help clarify these scenarios.

There are three DSFGs hosting a X-ray AGN (ADF22.A1, A4, A6) in Fig.~\ref{fig:md_n} (note that we excluded ADF22.A7, another AGN-host DSFG, in the analysis). The fact that both of ADF22.A1 and A4 host (heavily obscured) X-ray AGNs support the idea that growth of bulges and SMBHs are simultaneously accelerated in the most intense phase. With the limited samples in our hand, it is not straightforward to discuss the role of AGNs in quenching. 

In Fig.~\ref{fig:md_n}, ADF22.A11 occupies a unique parameter space. Its F444W S\'ersic index is close to the de Vaucouleurs profile, characteristic of local elliptical galaxies (\citealt{1948AnAp...11..247D}). ADF22.A11 could be in a transitional stage from DSFGs to quiescent galaxies (QGs), as also suggested by its position in the transitional rest-frame UVJ diagram (\citealt{2016MNRAS.455.3333K}; Kubo et al. in preparation).
On the other hand, significant dust emission is still detected. The dust emission profile is well described by an $n\simeq1$ model (Fig.~\ref{fig:almafit_n1}) with no signature of a dense dusty core. An explanation consistent with these observational clues is that inside-out quenching is occurring, where star formation activity has ceased in the innermost regions (resulting in the $n\sim4$ feature being visible only in the F444W profile, while the dust emission at very high resolution becomes faint), while star formation in the disk has not yet fully quenched. Such a transitional phase may have a short duration, with DSFGs generally following the evolutionary track suggested in Fig.~\ref{fig:md_n}, which could explain why only ADF22.A11 is identified in the ADF22 sample.

\subsection{Environment and evolution} \label{sec:discussion_environment}

The ADF22 field is situated in the heart of the SSA22 proto-cluster (\citealt{2015ApJ...815L...8U}), with the DSFGs ubiquitously embedded within the cosmic web filaments traced by Ly\,$\alpha$ emission (\citealt{2019Sci...366...97U}). In this section, we aim to connect the derived properties discussed above with this specific environment.

The NIRCam images, including F444W, reveal that the stellar mass distributions of DSFGs are generally characterized by a disk-like profile, similar to local late-type galaxies. The spatial locations of dust emission broadly align with those of the stellar components. These results indicate that DSFGs in the proto-cluster core are predominantly disky
galaxies (\citealt{2024arXiv241022155U}). This finding contrasts sharply with the well-established local morphology-density relation, where spheroidal galaxies dominate cluster centers while late-type disk galaxies are preferentially distributed in the surrounding, less dense environments (\citealt{1980ApJ...236..351D}).
Theoretical models suggest that phase space is conserved during the hierarchical collapse of structures, implying that the densest regions at high redshift will correspond to the densest regions in their descendants in the local universe. Hence, we are witnessing the origin of the morphology-density relation in the early universe, where the ancestors of local massive elliptical galaxies are undergoing a growth phase, rapidly assembling their mass as disk galaxies in proto-clusters. 
While this work sheds light on the situation for massive galaxies in the core of the proto-cluster, it presents an intriguing opportunity to further explore the outskirts of $z\sim3$ proto-clusters and/or less massive systems. Characterizing these regimes and redshift evolution will provide a more comprehensive understanding of galaxy evolution in proto-cluster environments.

To explain the high levels of star-formation activity observed in DSFGs, several scenarios have been proposed to effectively supply and consume gas fuel for this population, including cold gas accretion from the cosmic web (\citealt{2006MNRAS.368....2D}; \citealt{2009Natur.457..451D}) and major/minor mergers (e.g., \citealt{2006ApJS..163....1H}).
The morphologies revealed by NIRCam and ALMA images may support the scenario that the starbursts in the proto-cluster DSFGs are predominantly driven byminor perturbations or in-situ processes, such as disk instabilities (e.g., \citealt{2018Natur.560..613T}), possibly fueled by the abundant gas supply from the cosmic web (\citealt{2019Sci...366...97U}). For example, ADF22.A1 has a Toomre-$Q$ value below unity and exhibits a highly spun-up disk, consistent with an in-situ channel (\citealt{2024arXiv241022155U}). While the high density of DSFGs and neighboring galaxies suggests elevated rates of major and minor mergers, which likely play significant roles in their mass assembly, individual DSFGs in the starburst phase display morphologies reminiscent of late-type galaxies albeit disordered examples. This indicates that major mergers may not necessarily be the primary driver of starburst activity in these systems.

The configuration of ADF22.A7 is not straightforward to interpret. Photometric redshifts derived from non-resolved photometry support the idea that the system as a whole is at $z\sim3.09$ (\citealt{2014MNRAS.440.3462U}; \citealt{2015ApJ...815L...8U}), which is also consistent with the CO detections (\citealt{2019Sci...366...97U}; \citealt{2023A&A...679A.129R}; H. Umehata et al., in preparation). Meanwhile, the NIRCam images reveal two peaks, neither of which aligns with the dust peak. The NIRCam color image presents an extra red component at the dust position, though (Fig.~\ref{fig:nircam_alma}).
This suggests the possibility of a superposition of proto-cluster galaxies along the line of sight, 
Follow-up observations, such as those with JWST/NIRSpec-IFU, will help clarify the nature of this system.

\section{Conclusions}

We have presented high-resolution imaging of nine dusty star-forming galaxies (DSFGs) in the $z=3.1$ SSA22 proto-cluster, observed with both ALMA and JWST. These observations, reaching angular resolutions as fine as 0.05$^{\prime\prime}$ at 1.1\,mm, unveil the internal structures of DSFGs down to sub-kpc scales. The ALMA images reveal intricate spatial distributions of dust continuum emission, including compact dusty cores, elongated features suggestive of bars, and clumpy structures on scales down to $\sim350$\,pc. Meanwhile, JWST/NIRCam observations  provide unprecedented views of the stellar structures in these galaxies, also including three faint DSFGs in the vicinity of the bright ADF22.A4. For the first time, these observations reveal the internal stellar morphology and rest-frame optical-to-near-infrared colors of DSFGs in proto-clusters in detail.

The surface brightness distributions of six bright DSFGs were fit with S\'ersic profiles, utilizing ALMA 870\,$\mu$m and JWST F444W images. Both datasets suggest the existence of disk-like profiles in general, with S\'ersic indices ranging from $n\sim1$ to $n\sim2$ broadly. This range is consistent with a mixture of extended disks and central spheroidal components. The axis ratios, and position angles derived from the ALMA and JWST F444W data show good agreement, supporting the idea that dust continuum emission originates across the disk. The effective radii at 870\,$\mu$m tend to be smaller than those at F444W, plausibly influenced by active starburst actively occurred at the core.

Building on the growing body of evidence, including the ubiquitous detection of $n\sim1$ components and the strong correlation between the measured profile parameters of dust continuum and F444W, we argue that dust continuum emission originates across the disk. Furthermore, higher-resolution maps, reaching scales of down to 350\,pc, reveal intricate inner structures that are bright in dust continuum. In addition to central cores and clumps, offset ridges in dust continuum are suggested in some DSFGs, indicating significance of bars. 

Correlations between the dust mass surface density, stellar morphology, and core reddening (measured from rest-frame optical-to-near-infrared colors) reveal that the reddest cores are associated with the brightest and most compact dusty cores. This strongly suggests that dust extinction is the dominant factor responsible for the red colors observed in these DSFGs (see also \citealt{2024A&A...691A.299G}). Furthermore, DSFGs with dominant red cores often exhibit S\'ersic indices $n>1$, indicating the presence of possible stellar bulges. These observations may represent an accelerated phase of bulge growth in the early universe, possibly driven by bars. The diagram also indicates the transition from immature bulges to more settled, mature stages within DSFG populations.

In the core of the $z=3.1$ SSA22 proto-cluster, the DSFGs, associated with the cosmic web filaments, in general have a disk-like morphology. The DSFGs may assemble their mass mainly in a secular process, sustained by minor merger and gas accretion from the filaments. Notably, the most massive galaxies in the proto-cluster core tend to be late-type galaxies, presenting a striking contrast to the environmental dependence observed in the local universe. This stark difference suggests that we are witnessing the early stages of the formation of the local morphology-density relation.


\begin{acknowledgments}
This work is based on observations made with the NASA/ESA/CSA James Webb Space Telescope. The data were obtained from the Mikulski Archive for Space Telescopes at the Space Telescope Science Institute, which is operated by the Association of Universities for Research in Astronomy, Inc., under NASA contract NAS 5-03127 for JWST. These observations are associated with program \#3547.
This paper makes use of the following ALMA data: ADS/JAO.ALMA\#2019.1.00008.S,  2021.1.00071.S. ALMA is a partnership of ESO (representing its member states), NSF (USA) and NINS (Japan), together with NRC (Canada), NSTC and ASIAA (Taiwan), and KASI (Republic of Korea), in cooperation with the Republic of Chile. The Joint ALMA Observatory is operated by ESO, AUI/NRAO and NAOJ.
HU acknowledges support from JSPS KAKENHI Grant Numbers 20H01953, 22KK0231, 23K20240. 
This work was supported by NAOJ
ALMA Scientific Research Grant Numbers 2024-26A.
IRS acknowledge STFC support (ST/X001075/1).

\end{acknowledgments}

%

\vspace{5mm}


\software{{\sc astropy} \citep{2013A&A...558A..33A,2018AJ....156..123A}, {\sc casa}\citep{2022PASP..134k4501C}, {\sc carta}\citep{2021zndo...4905459C}, {\sc galfit}\citep{2010AJ....139.2097P}, {\sc lmfit}\citep{2016ascl.soft06014N},  {\sc matplotlib}\citep{2007CSE.....9...90H}, {\sc numpy}\citep{2020Natur.585..357H}, {\sc scipy}\citep{2020NatMe..17..261V}, 
          {\sc Source Extractor} \citep{1996A&AS..117..393B}
          }



\appendix

\section{Results from the profile fit with ALMA 870\,$\mu$m images}
\renewcommand{\thefigure}{A\arabic{figure}} 
\setcounter{figure}{0} 

\renewcommand{\thetable}{A\arabic{table}} 
\setcounter{table}{0} 

\begin{deluxetable}{lcccc}
\tabletypesize{\scriptsize}
\tablewidth{0pt} 
\tablecaption{ALMA 870\,$\mu$m profile measurements ($n=1$)\label{tab:ALMA_profilen1}}
\tablehead{
\colhead{ID} & \colhead{$n$} & \colhead{Re} & \colhead{b/a}  & \colhead{P.A.}\\
\colhead{} & \colhead{} & \colhead{[kpc]} & \colhead{}  & \colhead{[deg]}}
\startdata 
ADF22.A1 & 1.0 & 2.82 $\pm$ 0.31 & 0.38 $\pm$ 0.04 & 85.1 $\pm$ 8.6 \\
ADF22.A3 & 1.0 & 2.21 $\pm$ 0.23 & 0.16 $\pm$ 0.02 & 75.2 $\pm$ 7.5 \\
ADF22.A4 & 1.0 & 0.84 $\pm$ 0.08 & 0.89 $\pm$ 0.1 & 63.9 $\pm$ 13.9 \\
ADF22.A5 & 1.0 & 2.37 $\pm$ 0.23 & 0.21 $\pm$ 0.02 & 73.2 $\pm$ 7.3 \\
ADF22.A6 & 1.0 & 1.07 $\pm$ 0.15 & 0.64 $\pm$ 0.08 & 14.8 $\pm$ 4.1 \\
ADF22.A7 & 1.0 & 1.30 $\pm$ 0.15 & 0.42 $\pm$ 0.05 & 18.5 $\pm$ 2.4 \\
ADF22.A10 & 1.0 & 1.75 $\pm$ 0.23 & 0.29 $\pm$ 0.07 & 34.1 $\pm$ 4.8 \\
ADF22.A11 & 1.0 & 1.68 $\pm$ 0.38 & 0.4 $\pm$ 0.13 & 57.9 $\pm$ 9.6 \\
\enddata
\label{tab:almafit_n1}
\end{deluxetable}

\begin{figure*}
\epsscale{1.16}
\plotone{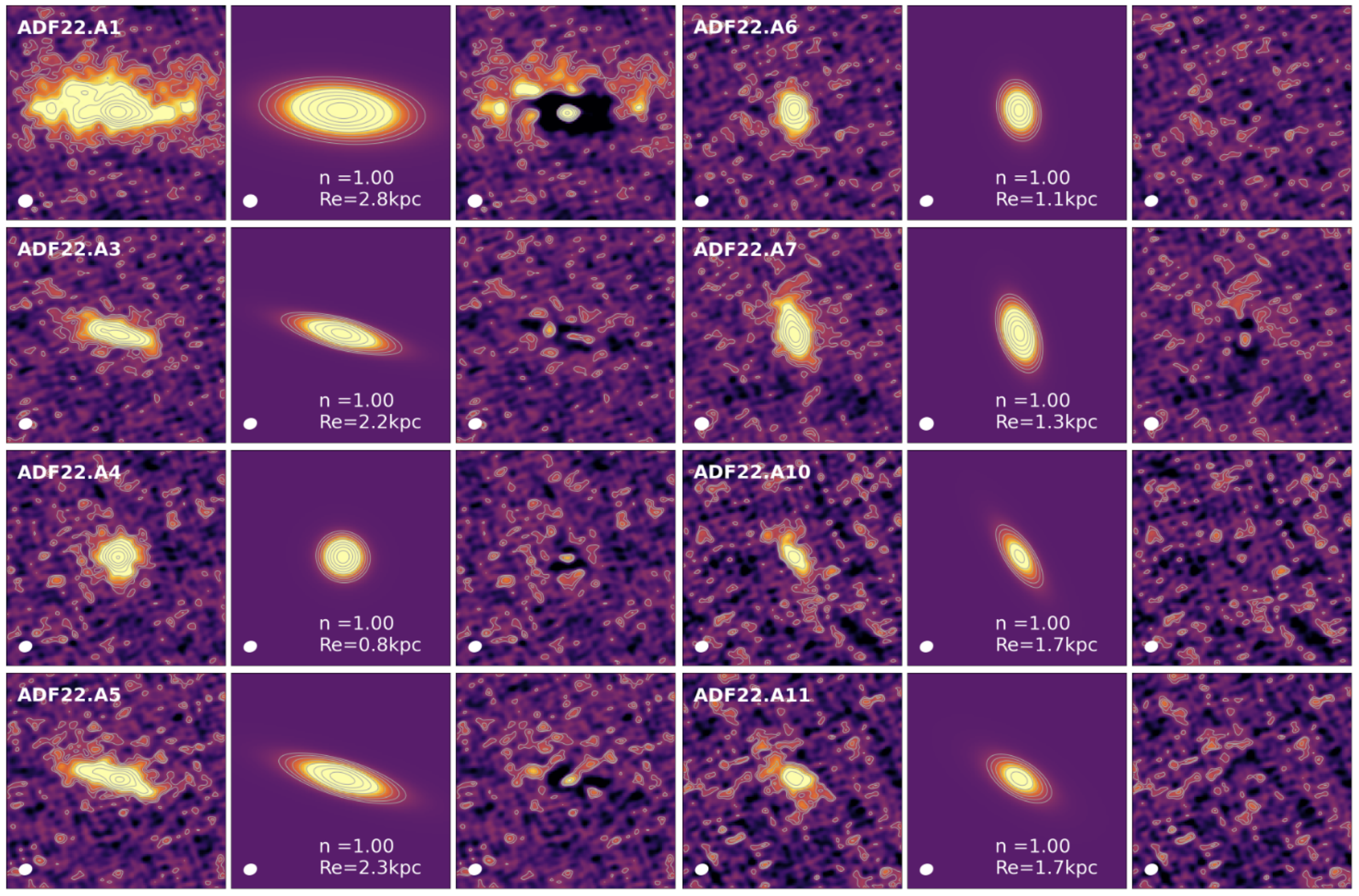}
\caption{
Results of the profile fit with ALMA 870\,$\mu$m imaging on the eight brightest DSFGs in ADF22 (0.15$^{\prime\prime}$ resolution) in the case of fixed $n$ ($n=1$) without bright emission masking as Fig.~\ref{fig:almafit}.
The fit generally causes positive and negative residuals, indicating a more complex profile for the bright DSFGs. The fits works better for the two faint DSFGs (ADF22.A10 and A11). We excluded ADF22.A16 since it is too faint for such an analysis.
}
\label{fig:almafit_n1}
\end{figure*}

\section{A close up view of ALMA and JWST images}
\renewcommand{\thefigure}{B\arabic{figure}} 
\setcounter{figure}{0} 

\renewcommand{\thetable}{B\arabic{table}} 
\setcounter{table}{0} 

\begin{figure}
\epsscale{0.75}
\plotone{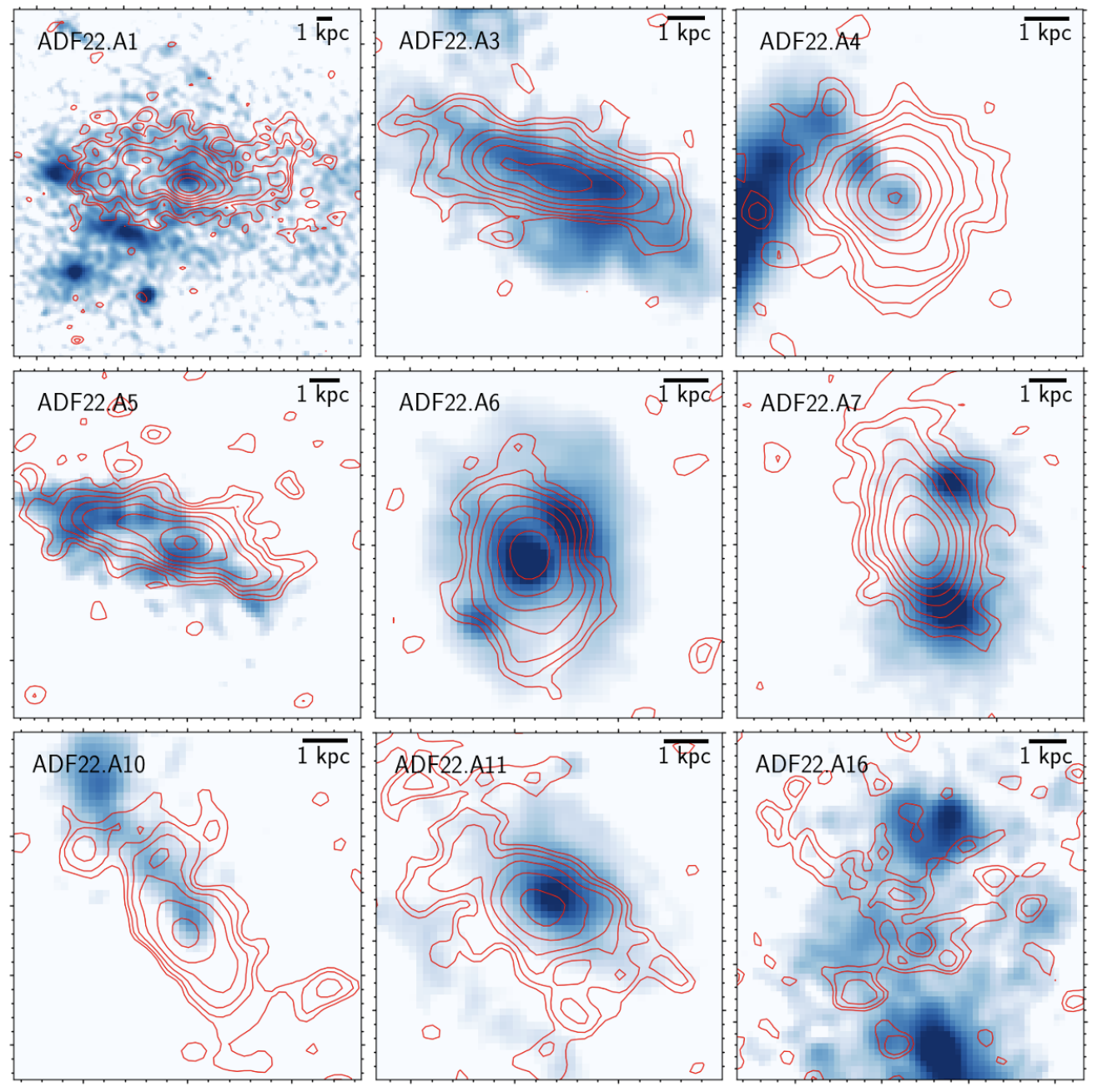}
\caption{
NIRCam F200W images of the nine DSFGs at original spatial resolution. Contours are $\sigma \times 1.5^n$ ($n=2-11$) of 870\,$\mu$m emission.
As shown, the relation between F200W (restframe $B$-band) emission and dust continuum emission is complex. The emission in the two bands is often segregated from each other, highlighting the significant dust obscuration at rest-frame $\approx$5000\AA. Some of DSFGs harbor compact clumps in the F200W image. Such {``blue-clumps''} can be attributed by young star clusters (or AGNs).
}
\label{fig:f200w_870}
\end{figure}


\bibliography{sample631.bbl}{}

\begin{thebibliography}{}
\expandafter\ifx\csname natexlab\endcsname\relax\def\natexlab#1{#1}\fi
\providecommand{\url}[1]{\href{#1}{#1}}
\providecommand{\dodoi}[1]{doi:~\href{http://doi.org/#1}{\nolinkurl{#1}}}
\providecommand{\doeprint}[1]{\href{http://ascl.net/#1}{\nolinkurl{http://ascl.net/#1}}}
\providecommand{\doarXiv}[1]{\href{https://arxiv.org/abs/#1}{\nolinkurl{https://arxiv.org/abs/#1}}}

\bibitem[{{Amvrosiadis} {et~al.}(2024){Amvrosiadis}, {Lange}, {Nightingale}, {He}, {Frenk}, {Oman}, {Smail}, {Swinbank}, {Fragkoudi}, {Gadotti}, {Cole}, {Borsato}, {Robertson}, {Massey}, {Cao}, \& {Li}}]{2024arXiv240401918A}
{Amvrosiadis}, A., {Lange}, S., {Nightingale}, J., {et~al.} 2024, arXiv e-prints, arXiv:2404.01918, \dodoi{10.48550/arXiv.2404.01918}

\bibitem[{{Astropy Collaboration} {et~al.}(2013){Astropy Collaboration}, {Robitaille}, {Tollerud}, {Greenfield}, {Droettboom}, {Bray}, {Aldcroft}, {Davis}, {Ginsburg}, {Price-Whelan}, {Kerzendorf}, {Conley}, {Crighton}, {Barbary}, {Muna}, {Ferguson}, {Grollier}, {Parikh}, {Nair}, {Unther}, {Deil}, {Woillez}, {Conseil}, {Kramer}, {Turner}, {Singer}, {Fox}, {Weaver}, {Zabalza}, {Edwards}, {Azalee Bostroem}, {Burke}, {Casey}, {Crawford}, {Dencheva}, {Ely}, {Jenness}, {Labrie}, {Lim}, {Pierfederici}, {Pontzen}, {Ptak}, {Refsdal}, {Servillat}, \& {Streicher}}]{2013A&A...558A..33A}
{Astropy Collaboration}, {Robitaille}, T.~P., {Tollerud}, E.~J., {et~al.} 2013, \aap, 558, A33, \dodoi{10.1051/0004-6361/201322068}

\bibitem[{{Astropy Collaboration} {et~al.}(2018){Astropy Collaboration}, {Price-Whelan}, {Sip{\H{o}}cz}, {G{\"u}nther}, {Lim}, {Crawford}, {Conseil}, {Shupe}, {Craig}, {Dencheva}, {Ginsburg}, {VanderPlas}, {Bradley}, {P{\'e}rez-Su{\'a}rez}, {de Val-Borro}, {Aldcroft}, {Cruz}, {Robitaille}, {Tollerud}, {Ardelean}, {Babej}, {Bach}, {Bachetti}, {Bakanov}, {Bamford}, {Barentsen}, {Barmby}, {Baumbach}, {Berry}, {Biscani}, {Boquien}, {Bostroem}, {Bouma}, {Brammer}, {Bray}, {Breytenbach}, {Buddelmeijer}, {Burke}, {Calderone}, {Cano Rodr{\'\i}guez}, {Cara}, {Cardoso}, {Cheedella}, {Copin}, {Corrales}, {Crichton}, {D'Avella}, {Deil}, {Depagne}, {Dietrich}, {Donath}, {Droettboom}, {Earl}, {Erben}, {Fabbro}, {Ferreira}, {Finethy}, {Fox}, {Garrison}, {Gibbons}, {Goldstein}, {Gommers}, {Greco}, {Greenfield}, {Groener}, {Grollier}, {Hagen}, {Hirst}, {Homeier}, {Horton}, {Hosseinzadeh}, {Hu}, {Hunkeler}, {Ivezi{\'c}}, {Jain}, {Jenness}, {Kanarek}, {Kendrew}, {Kern}, {Kerzendorf}, {Khvalko}, {King}, {Kirkby}, {Kulkarni},
  {Kumar}, {Lee}, {Lenz}, {Littlefair}, {Ma}, {Macleod}, {Mastropietro}, {McCully}, {Montagnac}, {Morris}, {Mueller}, {Mumford}, {Muna}, {Murphy}, {Nelson}, {Nguyen}, {Ninan}, {N{\"o}the}, {Ogaz}, {Oh}, {Parejko}, {Parley}, {Pascual}, {Patil}, {Patil}, {Plunkett}, {Prochaska}, {Rastogi}, {Reddy Janga}, {Sabater}, {Sakurikar}, {Seifert}, {Sherbert}, {Sherwood-Taylor}, {Shih}, {Sick}, {Silbiger}, {Singanamalla}, {Singer}, {Sladen}, {Sooley}, {Sornarajah}, {Streicher}, {Teuben}, {Thomas}, {Tremblay}, {Turner}, {Terr{\'o}n}, {van Kerkwijk}, {de la Vega}, {Watkins}, {Weaver}, {Whitmore}, {Woillez}, {Zabalza}, \& {Astropy Contributors}}]{2018AJ....156..123A}
{Astropy Collaboration}, {Price-Whelan}, A.~M., {Sip{\H{o}}cz}, B.~M., {et~al.} 2018, \aj, 156, 123, \dodoi{10.3847/1538-3881/aabc4f}

\bibitem[{{Bagley} {et~al.}(2023){Bagley}, {Finkelstein}, {Koekemoer}, {Ferguson}, {Arrabal Haro}, {Dickinson}, {Kartaltepe}, {Papovich}, {P{\'e}rez-Gonz{\'a}lez}, {Pirzkal}, {Somerville}, {Willmer}, {Yang}, {Yung}, {Fontana}, {Grazian}, {Grogin}, {Hirschmann}, {Kewley}, {Kirkpatrick}, {Kocevski}, {Lotz}, {Medrano}, {Morales}, {Pentericci}, {Ravindranath}, {Trump}, {Wilkins}, {Calabr{\`o}}, {Cooper}, {Costantin}, {de la Vega}, {Hilbert}, {Hutchison}, {Larson}, {Lucas}, {McGrath}, {Ryan}, {Wang}, \& {Wuyts}}]{2023ApJ...946L..12B}
{Bagley}, M.~B., {Finkelstein}, S.~L., {Koekemoer}, A.~M., {et~al.} 2023, \apjl, 946, L12, \dodoi{10.3847/2041-8213/acbb08}

\bibitem[{{Barger} {et~al.}(1998){Barger}, {Cowie}, {Sanders}, {Fulton}, {Taniguchi}, {Sato}, {Kawara}, \& {Okuda}}]{1998Natur.394..248B}
{Barger}, A.~J., {Cowie}, L.~L., {Sanders}, D.~B., {et~al.} 1998, \nat, 394, 248, \dodoi{10.1038/28338}

\bibitem[{{Bertin} \& {Arnouts}(1996)}]{1996A&AS..117..393B}
{Bertin}, E., \& {Arnouts}, S. 1996, \aaps, 117, 393, \dodoi{10.1051/aas:1996164}

\bibitem[{{Blain} {et~al.}(2004){Blain}, {Chapman}, {Smail}, \& {Ivison}}]{2004ApJ...611..725B}
{Blain}, A.~W., {Chapman}, S.~C., {Smail}, I., \& {Ivison}, R. 2004, \apj, 611, 725, \dodoi{10.1086/422353}

\bibitem[{{Blain} {et~al.}(2002){Blain}, {Smail}, {Ivison}, {Kneib}, \& {Frayer}}]{2002PhR...369..111B}
{Blain}, A.~W., {Smail}, I., {Ivison}, R.~J., {Kneib}, J.~P., \& {Frayer}, D.~T. 2002, \physrep, 369, 111, \dodoi{10.1016/S0370-1573(02)00134-5}

\bibitem[{{Bond} {et~al.}(1996){Bond}, {Kofman}, \& {Pogosyan}}]{1996Natur.380..603B}
{Bond}, J.~R., {Kofman}, L., \& {Pogosyan}, D. 1996, \nat, 380, 603, \dodoi{10.1038/380603a0}

\bibitem[{{Bournaud} {et~al.}(2008){Bournaud}, {Daddi}, {Elmegreen}, {Elmegreen}, {Nesvadba}, {Vanzella}, {Di Matteo}, {Le Tiran}, {Lehnert}, \& {Elbaz}}]{2008A&A...486..741B}
{Bournaud}, F., {Daddi}, E., {Elmegreen}, B.~G., {et~al.} 2008, \aap, 486, 741, \dodoi{10.1051/0004-6361:20079250}

\bibitem[{{CASA Team} {et~al.}(2022){CASA Team}, {Bean}, {Bhatnagar}, {Castro}, {Donovan Meyer}, {Emonts}, {Garcia}, {Garwood}, {Golap}, {Gonzalez Villalba}, {Harris}, {Hayashi}, {Hoskins}, {Hsieh}, {Jagannathan}, {Kawasaki}, {Keimpema}, {Kettenis}, {Lopez}, {Marvil}, {Masters}, {McNichols}, {Mehringer}, {Miel}, {Moellenbrock}, {Montesino}, {Nakazato}, {Ott}, {Petry}, {Pokorny}, {Raba}, {Rau}, {Schiebel}, {Schweighart}, {Sekhar}, {Shimada}, {Small}, {Steeb}, {Sugimoto}, {Suoranta}, {Tsutsumi}, {van Bemmel}, {Verkouter}, {Wells}, {Xiong}, {Szomoru}, {Griffith}, {Glendenning}, \& {Kern}}]{2022PASP..134k4501C}
{CASA Team}, {Bean}, B., {Bhatnagar}, S., {et~al.} 2022, \pasp, 134, 114501, \dodoi{10.1088/1538-3873/ac9642}

\bibitem[{{Casey} {et~al.}(2014){Casey}, {Narayanan}, \& {Cooray}}]{2014PhR...541...45C}
{Casey}, C.~M., {Narayanan}, D., \& {Cooray}, A. 2014, \physrep, 541, 45, \dodoi{10.1016/j.physrep.2014.02.009}

\bibitem[{{Chen} {et~al.}(2015){Chen}, {Smail}, {Swinbank}, {Simpson}, {Ma}, {Alexander}, {Biggs}, {Brandt}, {Chapman}, {Coppin}, {Danielson}, {Dannerbauer}, {Edge}, {Greve}, {Ivison}, {Karim}, {Menten}, {Schinnerer}, {Walter}, {Wardlow}, {Wei{\ss}}, \& {van der Werf}}]{2015ApJ...799..194C}
{Chen}, C.-C., {Smail}, I., {Swinbank}, A.~M., {et~al.} 2015, \apj, 799, 194, \dodoi{10.1088/0004-637X/799/2/194}

\bibitem[{{Chen} {et~al.}(2022){Chen}, {Gao}, {Hsu}, {Liao}, {Ling}, {Lo}, {Smail}, {Wang}, \& {Wang}}]{2022ApJ...939L...7C}
{Chen}, C.-C., {Gao}, Z.-K., {Hsu}, Q.-N., {et~al.} 2022, \apjl, 939, L7, \dodoi{10.3847/2041-8213/ac98c6}

\bibitem[{{Cheng} {et~al.}(2023){Cheng}, {Huang}, {Smail}, {Yan}, {Cohen}, {Jansen}, {Windhorst}, {Ma}, {Koekemoer}, {Willmer}, {Willner}, {Diego}, {Frye}, {Conselice}, {Ferreira}, {Petric}, {Yun}, {Gim}, {Polletta}, {Duncan}, {Holwerda}, {R{\"o}ttgering}, {Honor}, {Hathi}, {Kamieneski}, {Adams}, {Coe}, {Broadhurst}, {Summers}, {Tompkins}, {Driver}, {Grogin}, {Marshall}, {Pirzkal}, {Robotham}, \& {Ryan}}]{2023ApJ...942L..19C}
{Cheng}, C., {Huang}, J.-S., {Smail}, I., {et~al.} 2023, \apjl, 942, L19, \dodoi{10.3847/2041-8213/aca9d0}

\bibitem[{{Colina} {et~al.}(2023){Colina}, {Crespo G{\'o}mez}, {{\'A}lvarez-M{\'a}rquez}, {Bik}, {Walter}, {Boogaard}, {Labiano}, {Peissker}, {P{\'e}rez-Gonz{\'a}lez}, {{\"O}stlin}, {Greve}, {N{\o}rgaard-Nielsen}, {Wright}, {Alonso-Herrero}, {Azollini}, {Caputi}, {Dicken}, {Garc{\'\i}a-Mar{\'\i}n}, {Hjorth}, {Ilbert}, {Kendrew}, {Pye}, {Tikkanen}, {van der Werf}, {Costantin}, {Iani}, {Gillman}, {Jermann}, {Langeroodi}, {Moutard}, {Rinaldi}, {Topinka}, {van Dishoeck}, {G{\"u}del}, {Henning}, {Lagage}, {Ray}, \& {Vandenbussche}}]{2023A&A...673L...6C}
{Colina}, L., {Crespo G{\'o}mez}, A., {{\'A}lvarez-M{\'a}rquez}, J., {et~al.} 2023, \aap, 673, L6, \dodoi{10.1051/0004-6361/202346535}

\bibitem[{{Comrie} {et~al.}(2021){Comrie}, {Wang}, {Hsu}, {Moraghan}, {Harris}, {Pang}, {Pi{\'n}ska}, {Chiang}, {Chang}, {Hwang}, {Jan}, {Lin}, \& {Simmonds}}]{2021zndo...4905459C}
{Comrie}, A., {Wang}, K.-S., {Hsu}, S.-C., {et~al.} 2021, {CARTA: The Cube Analysis and Rendering Tool for Astronomy}, 2.0.0,  Zenodo, \dodoi{10.5281/zenodo.4905459}

\bibitem[{{Conselice} {et~al.}(2004){Conselice}, {Grogin}, {Jogee}, {Lucas}, {Dahlen}, {de Mello}, {Gardner}, {Mobasher}, \& {Ravindranath}}]{2004ApJ...600L.139C}
{Conselice}, C.~J., {Grogin}, N.~A., {Jogee}, S., {et~al.} 2004, \apjl, 600, L139, \dodoi{10.1086/378556}

\bibitem[{{da Cunha} {et~al.}(2013){da Cunha}, {Groves}, {Walter}, {Decarli}, {Weiss}, {Bertoldi}, {Carilli}, {Daddi}, {Elbaz}, {Ivison}, {Maiolino}, {Riechers}, {Rix}, {Sargent}, \& {Smail}}]{2013ApJ...766...13D}
{da Cunha}, E., {Groves}, B., {Walter}, F., {et~al.} 2013, \apj, 766, 13, \dodoi{10.1088/0004-637X/766/1/13}

\bibitem[{{Daddi} {et~al.}(2009){Daddi}, {Dannerbauer}, {Stern}, {Dickinson}, {Morrison}, {Elbaz}, {Giavalisco}, {Mancini}, {Pope}, \& {Spinrad}}]{2009ApJ...694.1517D}
{Daddi}, E., {Dannerbauer}, H., {Stern}, D., {et~al.} 2009, \apj, 694, 1517, \dodoi{10.1088/0004-637X/694/2/1517}

\bibitem[{{De Lucia} {et~al.}(2006){De Lucia}, {Springel}, {White}, {Croton}, \& {Kauffmann}}]{2006MNRAS.366..499D}
{De Lucia}, G., {Springel}, V., {White}, S. D.~M., {Croton}, D., \& {Kauffmann}, G. 2006, \mnras, 366, 499, \dodoi{10.1111/j.1365-2966.2005.09879.x}

\bibitem[{{de Vaucouleurs}(1948)}]{1948AnAp...11..247D}
{de Vaucouleurs}, G. 1948, Annales d'Astrophysique, 11, 247

\bibitem[{{Dekel} \& {Birnboim}(2006)}]{2006MNRAS.368....2D}
{Dekel}, A., \& {Birnboim}, Y. 2006, \mnras, 368, 2, \dodoi{10.1111/j.1365-2966.2006.10145.x}

\bibitem[{{Dekel} {et~al.}(2009){Dekel}, {Birnboim}, {Engel}, {Freundlich}, {Goerdt}, {Mumcuoglu}, {Neistein}, {Pichon}, {Teyssier}, \& {Zinger}}]{2009Natur.457..451D}
{Dekel}, A., {Birnboim}, Y., {Engel}, G., {et~al.} 2009, \nat, 457, 451, \dodoi{10.1038/nature07648}

\bibitem[{{Dressler}(1980)}]{1980ApJ...236..351D}
{Dressler}, A. 1980, \apj, 236, 351, \dodoi{10.1086/157753}

\bibitem[{{Dunlop} {et~al.}(2017){Dunlop}, {McLure}, {Biggs}, {Geach}, {Micha{\l}owski}, {Ivison}, {Rujopakarn}, {van Kampen}, {Kirkpatrick}, {Pope}, {Scott}, {Swinbank}, {Targett}, {Aretxaga}, {Austermann}, {Best}, {Bruce}, {Chapin}, {Charlot}, {Cirasuolo}, {Coppin}, {Ellis}, {Finkelstein}, {Hayward}, {Hughes}, {Ibar}, {Jagannathan}, {Khochfar}, {Koprowski}, {Narayanan}, {Nyland}, {Papovich}, {Peacock}, {Rieke}, {Robertson}, {Vernstrom}, {Werf}, {Wilson}, \& {Yun}}]{2017MNRAS.466..861D}
{Dunlop}, J.~S., {McLure}, R.~J., {Biggs}, A.~D., {et~al.} 2017, \mnras, 466, 861, \dodoi{10.1093/mnras/stw3088}

\bibitem[{{Eales} {et~al.}(1999){Eales}, {Lilly}, {Gear}, {Dunne}, {Bond}, {Hammer}, {Le F{\`e}vre}, \& {Crampton}}]{1999ApJ...515..518E}
{Eales}, S., {Lilly}, S., {Gear}, W., {et~al.} 1999, \apj, 515, 518, \dodoi{10.1086/307069}

\bibitem[{{Ellis} {et~al.}(1997){Ellis}, {Smail}, {Dressler}, {Couch}, {Oemler}, {Butcher}, \& {Sharples}}]{1997ApJ...483..582E}
{Ellis}, R.~S., {Smail}, I., {Dressler}, A., {et~al.} 1997, \apj, 483, 582, \dodoi{10.1086/304261}

\bibitem[{{Elmegreen} \& {Elmegreen}(2005)}]{2005ApJ...627..632E}
{Elmegreen}, B.~G., \& {Elmegreen}, D.~M. 2005, \apj, 627, 632, \dodoi{10.1086/430514}

\bibitem[{{Elmegreen} {et~al.}(2009){Elmegreen}, {Elmegreen}, {Fernandez}, \& {Lemonias}}]{2009ApJ...692...12E}
{Elmegreen}, B.~G., {Elmegreen}, D.~M., {Fernandez}, M.~X., \& {Lemonias}, J.~J. 2009, \apj, 692, 12, \dodoi{10.1088/0004-637X/692/1/12}

\bibitem[{{Elmegreen} {et~al.}(2007){Elmegreen}, {Elmegreen}, {Ravindranath}, \& {Coe}}]{2007ApJ...658..763E}
{Elmegreen}, D.~M., {Elmegreen}, B.~G., {Ravindranath}, S., \& {Coe}, D.~A. 2007, \apj, 658, 763, \dodoi{10.1086/511667}

\bibitem[{{Genzel} {et~al.}(2008){Genzel}, {Burkert}, {Bouch{\'e}}, {Cresci}, {F{\"o}rster Schreiber}, {Shapley}, {Shapiro}, {Tacconi}, {Buschkamp}, {Cimatti}, {Daddi}, {Davies}, {Eisenhauer}, {Erb}, {Genel}, {Gerhard}, {Hicks}, {Lutz}, {Naab}, {Ott}, {Rabien}, {Renzini}, {Steidel}, {Sternberg}, \& {Lilly}}]{2008ApJ...687...59G}
{Genzel}, R., {Burkert}, A., {Bouch{\'e}}, N., {et~al.} 2008, \apj, 687, 59, \dodoi{10.1086/591840}

\bibitem[{{Genzel} {et~al.}(2011){Genzel}, {Newman}, {Jones}, {F{\"o}rster Schreiber}, {Shapiro}, {Genel}, {Lilly}, {Renzini}, {Tacconi}, {Bouch{\'e}}, {Burkert}, {Cresci}, {Buschkamp}, {Carollo}, {Ceverino}, {Davies}, {Dekel}, {Eisenhauer}, {Hicks}, {Kurk}, {Lutz}, {Mancini}, {Naab}, {Peng}, {Sternberg}, {Vergani}, \& {Zamorani}}]{2011ApJ...733..101G}
{Genzel}, R., {Newman}, S., {Jones}, T., {et~al.} 2011, \apj, 733, 101, \dodoi{10.1088/0004-637X/733/2/101}

\bibitem[{{Gillman} {et~al.}(2023){Gillman}, {Gullberg}, {Brammer}, {Vijayan}, {Lee}, {Bl{\'a}nquez}, {Brinch}, {Greve}, {Jermann}, {Jin}, {Kokorev}, {Liu}, {Magdis}, {Rizzo}, \& {Valentino}}]{2023A&A...676A..26G}
{Gillman}, S., {Gullberg}, B., {Brammer}, G., {et~al.} 2023, \aap, 676, A26, \dodoi{10.1051/0004-6361/202346531}

\bibitem[{{Gillman} {et~al.}(2024){Gillman}, {Smail}, {Gullberg}, {Swinbank}, {Vijayan}, {Lee}, {Brammer}, {Dudzevi{\v{c}}i{\={u}}t{\.{e}}}, {Greve}, {Almaini}, {Brinch}, {Chapman}, {Chen}, {Ikarashi}, {Matsuda}, {Wang}, {Walter}, \& {van der Werf}}]{2024A&A...691A.299G}
{Gillman}, S., {Smail}, I., {Gullberg}, B., {et~al.} 2024, \aap, 691, A299, \dodoi{10.1051/0004-6361/202451006}

\bibitem[{{Greve} {et~al.}(2005){Greve}, {Bertoldi}, {Smail}, {Neri}, {Chapman}, {Blain}, {Ivison}, {Genzel}, {Omont}, {Cox}, {Tacconi}, \& {Kneib}}]{2005MNRAS.359.1165G}
{Greve}, T.~R., {Bertoldi}, F., {Smail}, I., {et~al.} 2005, \mnras, 359, 1165, \dodoi{10.1111/j.1365-2966.2005.08979.x}

\bibitem[{{Gullberg} {et~al.}(2018){Gullberg}, {Swinbank}, {Smail}, {Biggs}, {Bertoldi}, {De Breuck}, {Chapman}, {Chen}, {Cooke}, {Coppin}, {Cox}, {Dannerbauer}, {Dunlop}, {Edge}, {Farrah}, {Geach}, {Greve}, {Hodge}, {Ibar}, {Ivison}, {Karim}, {Schinnerer}, {Scott}, {Simpson}, {Stach}, {Thomson}, {van der Werf}, {Walter}, {Wardlow}, \& {Weiss}}]{2018ApJ...859...12G}
{Gullberg}, B., {Swinbank}, A.~M., {Smail}, I., {et~al.} 2018, \apj, 859, 12, \dodoi{10.3847/1538-4357/aabe8c}

\bibitem[{{Gullberg} {et~al.}(2019){Gullberg}, {Smail}, {Swinbank}, {Dudzevi{\v{c}}i{\={u}}t{\.{e}}}, {Stach}, {Thomson}, {Almaini}, {Chen}, {Conselice}, {Cooke}, {Farrah}, {Ivison}, {Maltby}, {Micha{\l}owski}, {Simpson}, {Scott}, {Wardlow}, \& {Weiss}}]{2019MNRAS.490.4956G}
{Gullberg}, B., {Smail}, I., {Swinbank}, A.~M., {et~al.} 2019, \mnras, 490, 4956, \dodoi{10.1093/mnras/stz2835}

\bibitem[{{Hainline} {et~al.}(2011){Hainline}, {Blain}, {Smail}, {Alexander}, {Armus}, {Chapman}, \& {Ivison}}]{2011ApJ...740...96H}
{Hainline}, L.~J., {Blain}, A.~W., {Smail}, I., {et~al.} 2011, \apj, 740, 96, \dodoi{10.1088/0004-637X/740/2/96}

\bibitem[{{Harris} {et~al.}(2020){Harris}, {Millman}, {van der Walt}, {Gommers}, {Virtanen}, {Cournapeau}, {Wieser}, {Taylor}, {Berg}, {Smith}, {Kern}, {Picus}, {Hoyer}, {van Kerkwijk}, {Brett}, {Haldane}, {del R{\'\i}o}, {Wiebe}, {Peterson}, {G{\'e}rard-Marchant}, {Sheppard}, {Reddy}, {Weckesser}, {Abbasi}, {Gohlke}, \& {Oliphant}}]{2020Natur.585..357H}
{Harris}, C.~R., {Millman}, K.~J., {van der Walt}, S.~J., {et~al.} 2020, \nat, 585, 357, \dodoi{10.1038/s41586-020-2649-2}

\bibitem[{{Hayashino} {et~al.}(2004){Hayashino}, {Matsuda}, {Tamura}, {Yamauchi}, {Yamada}, {Ajiki}, {Fujita}, {Murayama}, {Nagao}, {Ohta}, {Okamura}, {Ouchi}, {Shimasaku}, {Shioya}, \& {Taniguchi}}]{2004AJ....128.2073H}
{Hayashino}, T., {Matsuda}, Y., {Tamura}, H., {et~al.} 2004, \aj, 128, 2073, \dodoi{10.1086/424935}

\bibitem[{{Hayatsu} {et~al.}(2017){Hayatsu}, {Matsuda}, {Umehata}, {Yoshida}, {Smail}, {Swinbank}, {Ivison}, {Kohno}, {Tamura}, {Kubo}, {Iono}, {Hatsukade}, {Nakanishi}, {Kawabe}, {Nagao}, {Inoue}, {Takeuchi}, {Lee}, {Ao}, {Fujimoto}, {Izumi}, {Yamaguchi}, {Ikarashi}, \& {Yamada}}]{2017PASJ...69...45H}
{Hayatsu}, N.~H., {Matsuda}, Y., {Umehata}, H., {et~al.} 2017, \pasj, 69, 45, \dodoi{10.1093/pasj/psx018}

\bibitem[{{Hodge} \& {da Cunha}(2020)}]{2020RSOS....700556H}
{Hodge}, J.~A., \& {da Cunha}, E. 2020, Royal Society Open Science, 7, 200556, \dodoi{10.1098/rsos.200556}

\bibitem[{{Hodge} {et~al.}(2013){Hodge}, {Karim}, {Smail}, {Swinbank}, {Walter}, {Biggs}, {Ivison}, {Weiss}, {Alexander}, {Bertoldi}, {Brandt}, {Chapman}, {Coppin}, {Cox}, {Danielson}, {Dannerbauer}, {De Breuck}, {Decarli}, {Edge}, {Greve}, {Knudsen}, {Menten}, {Rix}, {Schinnerer}, {Simpson}, {Wardlow}, \& {van der Werf}}]{2013ApJ...768...91H}
{Hodge}, J.~A., {Karim}, A., {Smail}, I., {et~al.} 2013, \apj, 768, 91, \dodoi{10.1088/0004-637X/768/1/91}

\bibitem[{{Hodge} {et~al.}(2016){Hodge}, {Swinbank}, {Simpson}, {Smail}, {Walter}, {Alexander}, {Bertoldi}, {Biggs}, {Brandt}, {Chapman}, {Chen}, {Coppin}, {Cox}, {Dannerbauer}, {Edge}, {Greve}, {Ivison}, {Karim}, {Knudsen}, {Menten}, {Rix}, {Schinnerer}, {Wardlow}, {Weiss}, \& {van der Werf}}]{2016ApJ...833..103H}
{Hodge}, J.~A., {Swinbank}, A.~M., {Simpson}, J.~M., {et~al.} 2016, \apj, 833, 103, \dodoi{10.3847/1538-4357/833/1/103}

\bibitem[{{Hodge} {et~al.}(2019){Hodge}, {Smail}, {Walter}, {da Cunha}, {Swinbank}, {Rybak}, {Venemans}, {Brandt}, {Calistro Rivera}, {Chapman}, {Chen}, {Cox}, {Dannerbauer}, {Decarli}, {Greve}, {Knudsen}, {Menten}, {Schinnerer}, {Simpson}, {van der Werf}, {Wardlow}, \& {Weiss}}]{2019ApJ...876..130H}
{Hodge}, J.~A., {Smail}, I., {Walter}, F., {et~al.} 2019, \apj, 876, 130, \dodoi{10.3847/1538-4357/ab1846}

\bibitem[{{Hodge} {et~al.}(2025){Hodge}, {Cunha}, {Kendrew}, {Li}, {Smail}, {Westoby}, {Nayak}, {Swinbank}, {Chen}, {Walter}, {van der Werf}, {Cracraft}, {Battisti}, {Brandt}, {Calistro Rivera}, {Chapman}, {Cox}, {Dannerbauer}, {Decarli}, {Frias Castillo}, {Greve}, {Knudsen}, {Leslie}, {Menten}, {Rybak}, {Schinnerer}, {Wardlow}, \& {Weiss}}]{2025ApJ...978..165H}
{Hodge}, J.~A., {Cunha}, E.~d., {Kendrew}, S., {et~al.} 2025, \apj, 978, 165, \dodoi{10.3847/1538-4357/ad9a52}

\bibitem[{{Hopkins} {et~al.}(2006){Hopkins}, {Hernquist}, {Cox}, {Di Matteo}, {Robertson}, \& {Springel}}]{2006ApJS..163....1H}
{Hopkins}, P.~F., {Hernquist}, L., {Cox}, T.~J., {et~al.} 2006, \apjs, 163, 1, \dodoi{10.1086/499298}

\bibitem[{{Hughes} {et~al.}(1998){Hughes}, {Serjeant}, {Dunlop}, {Rowan-Robinson}, {Blain}, {Mann}, {Ivison}, {Peacock}, {Efstathiou}, {Gear}, {Oliver}, {Lawrence}, {Longair}, {Goldschmidt}, \& {Jenness}}]{1998Natur.394..241H}
{Hughes}, D.~H., {Serjeant}, S., {Dunlop}, J., {et~al.} 1998, \nat, 394, 241, \dodoi{10.1038/28328}

\bibitem[{{Hunter}(2007)}]{2007CSE.....9...90H}
{Hunter}, J.~D. 2007, Computing in Science and Engineering, 9, 90, \dodoi{10.1109/MCSE.2007.55}

\bibitem[{{Ikarashi} {et~al.}(2022){Ikarashi}, {Ivison}, {Cowley}, \& {Kohno}}]{2022A&A...659A.154I}
{Ikarashi}, S., {Ivison}, R.~J., {Cowley}, W.~I., \& {Kohno}, K. 2022, \aap, 659, A154, \dodoi{10.1051/0004-6361/202141196}

\bibitem[{{Ikarashi} {et~al.}(2015){Ikarashi}, {Ivison}, {Caputi}, {Aretxaga}, {Dunlop}, {Hatsukade}, {Hughes}, {Iono}, {Izumi}, {Kawabe}, {Kohno}, {Lagos}, {Motohara}, {Nakanishi}, {Ohta}, {Tamura}, {Umehata}, {Wilson}, {Yabe}, \& {Yun}}]{2015ApJ...810..133I}
{Ikarashi}, S., {Ivison}, R.~J., {Caputi}, K.~I., {et~al.} 2015, \apj, 810, 133, \dodoi{10.1088/0004-637X/810/2/133}

\bibitem[{{Iono} {et~al.}(2016){Iono}, {Yun}, {Aretxaga}, {Hatsukade}, {Hughes}, {Ikarashi}, {Izumi}, {Kawabe}, {Kohno}, {Lee}, {Matsuda}, {Nakanishi}, {Saito}, {Tamura}, {Ueda}, {Umehata}, {Wilson}, {Michiyama}, \& {Ando}}]{2016ApJ...829L..10I}
{Iono}, D., {Yun}, M.~S., {Aretxaga}, I., {et~al.} 2016, \apjl, 829, L10, \dodoi{10.3847/2041-8205/829/1/L10}

\bibitem[{{Ivison} {et~al.}(1998){Ivison}, {Smail}, {Le Borgne}, {Blain}, {Kneib}, {Bezecourt}, {Kerr}, \& {Davies}}]{1998MNRAS.298..583I}
{Ivison}, R.~J., {Smail}, I., {Le Borgne}, J.~F., {et~al.} 1998, \mnras, 298, 583, \dodoi{10.1046/j.1365-8711.1998.01677.x}

\bibitem[{{Kodama} {et~al.}(1998){Kodama}, {Arimoto}, {Barger}, \& {Arag'on-Salamanca}}]{1998A&A...334...99K}
{Kodama}, T., {Arimoto}, N., {Barger}, A.~J., \& {Arag'on-Salamanca}, A. 1998, \aap, 334, 99, \dodoi{10.48550/arXiv.astro-ph/9802245}

\bibitem[{{Kodama} {et~al.}(2007){Kodama}, {Tanaka}, {Kajisawa}, {Kurk}, {Venemans}, {De Breuck}, {Vernet}, \& {Lidman}}]{2007MNRAS.377.1717K}
{Kodama}, T., {Tanaka}, I., {Kajisawa}, M., {et~al.} 2007, \mnras, 377, 1717, \dodoi{10.1111/j.1365-2966.2007.11739.x}

\bibitem[{{Kruk} {et~al.}(2018){Kruk}, {Lintott}, {Bamford}, {Masters}, {Simmons}, {H{\"a}u{\ss}ler}, {Cardamone}, {Hart}, {Kelvin}, {Schawinski}, {Smethurst}, \& {Vika}}]{2018MNRAS.473.4731K}
{Kruk}, S.~J., {Lintott}, C.~J., {Bamford}, S.~P., {et~al.} 2018, \mnras, 473, 4731, \dodoi{10.1093/mnras/stx2605}

\bibitem[{{Kubo} {et~al.}(2016){Kubo}, {Yamada}, {Ichikawa}, {Kajisawa}, {Matsuda}, {Tanaka}, \& {Umehata}}]{2016MNRAS.455.3333K}
{Kubo}, M., {Yamada}, T., {Ichikawa}, T., {et~al.} 2016, \mnras, 455, 3333, \dodoi{10.1093/mnras/stv2392}

\bibitem[{{Lemson} \& {Kauffmann}(1999)}]{1999MNRAS.302..111L}
{Lemson}, G., \& {Kauffmann}, G. 1999, \mnras, 302, 111, \dodoi{10.1046/j.1365-8711.1999.02090.x}

\bibitem[{{Li} \& {Draine}(2001)}]{2001ApJ...554..778L}
{Li}, A., \& {Draine}, B.~T. 2001, \apj, 554, 778, \dodoi{10.1086/323147}

\bibitem[{{Matsuda} {et~al.}(2005){Matsuda}, {Yamada}, {Hayashino}, {Tamura}, {Yamauchi}, {Murayama}, {Nagao}, {Ohta}, {Okamura}, {Ouchi}, {Shimasaku}, {Shioya}, \& {Taniguchi}}]{2005ApJ...634L.125M}
{Matsuda}, Y., {Yamada}, T., {Hayashino}, T., {et~al.} 2005, \apjl, 634, L125, \dodoi{10.1086/499071}

\bibitem[{{Miller} {et~al.}(2018){Miller}, {Chapman}, {Aravena}, {Ashby}, {Hayward}, {Vieira}, {Wei{\ss}}, {Babul}, {B{\'e}thermin}, {Bradford}, {Brodwin}, {Carlstrom}, {Chen}, {Cunningham}, {De Breuck}, {Gonzalez}, {Greve}, {Harnett}, {Hezaveh}, {Lacaille}, {Litke}, {Ma}, {Malkan}, {Marrone}, {Morningstar}, {Murphy}, {Narayanan}, {Pass}, {Perry}, {Phadke}, {Rennehan}, {Rotermund}, {Simpson}, {Spilker}, {Sreevani}, {Stark}, {Strandet}, \& {Strom}}]{2018Natur.556..469M}
{Miller}, T.~B., {Chapman}, S.~C., {Aravena}, M., {et~al.} 2018, \nat, 556, 469, \dodoi{10.1038/s41586-018-0025-2}

\bibitem[{{Monson} {et~al.}(2021){Monson}, {Lehmer}, {Doore}, {Eufrasio}, {Bonine}, {Alexander}, {Harrison}, {Kubo}, {Mantha}, {Saez}, {Straughn}, \& {Umehata}}]{2021ApJ...919...51M}
{Monson}, E.~B., {Lehmer}, B.~D., {Doore}, K., {et~al.} 2021, \apj, 919, 51, \dodoi{10.3847/1538-4357/ac0f84}

\bibitem[{{Monson} {et~al.}(2023){Monson}, {Doore}, {Eufrasio}, {Lehmer}, {Alexander}, {Harrison}, {Kubo}, {Saez}, \& {Umehata}}]{2023ApJ...951...15M}
{Monson}, E.~B., {Doore}, K., {Eufrasio}, R.~T., {et~al.} 2023, \apj, 951, 15, \dodoi{10.3847/1538-4357/acd449}

\bibitem[{{Newville} {et~al.}(2016){Newville}, {Stensitzki}, {Allen}, {Rawlik}, {Ingargiola}, \& {Nelson}}]{2016ascl.soft06014N}
{Newville}, M., {Stensitzki}, T., {Allen}, D.~B., {et~al.} 2016, {Lmfit: Non-Linear Least-Square Minimization and Curve-Fitting for Python}, Astrophysics Source Code Library, record ascl:1606.014

\bibitem[{{Oteo} {et~al.}(2018){Oteo}, {Ivison}, {Dunne}, {Manilla-Robles}, {Maddox}, {Lewis}, {de Zotti}, {Bremer}, {Clements}, {Cooray}, {Dannerbauer}, {Eales}, {Greenslade}, {Omont}, {Perez{\textendash}Fourn{\'o}n}, {Riechers}, {Scott}, {van der Werf}, {Weiss}, \& {Zhang}}]{2018ApJ...856...72O}
{Oteo}, I., {Ivison}, R.~J., {Dunne}, L., {et~al.} 2018, \apj, 856, 72, \dodoi{10.3847/1538-4357/aaa1f1}

\bibitem[{{Peng} {et~al.}(2002){Peng}, {Ho}, {Impey}, \& {Rix}}]{2002AJ....124..266P}
{Peng}, C.~Y., {Ho}, L.~C., {Impey}, C.~D., \& {Rix}, H.-W. 2002, \aj, 124, 266, \dodoi{10.1086/340952}

\bibitem[{{Peng} {et~al.}(2010){Peng}, {Ho}, {Impey}, \& {Rix}}]{2010AJ....139.2097P}
---. 2010, \aj, 139, 2097, \dodoi{10.1088/0004-6256/139/6/2097}

\bibitem[{{Perrin} {et~al.}(2014){Perrin}, {Sivaramakrishnan}, {Lajoie}, {Elliott}, {Pueyo}, {Ravindranath}, \& {Albert}}]{2014SPIE.9143E..3XP}
{Perrin}, M.~D., {Sivaramakrishnan}, A., {Lajoie}, C.-P., {et~al.} 2014, in Society of Photo-Optical Instrumentation Engineers (SPIE) Conference Series, Vol. 9143, Space Telescopes and Instrumentation 2014: Optical, Infrared, and Millimeter Wave, ed. J.~{Oschmann}, Jacobus~M., M.~{Clampin}, G.~G. {Fazio}, \& H.~A. {MacEwen}, 91433X, \dodoi{10.1117/12.2056689}

\bibitem[{{Popping} {et~al.}(2022){Popping}, {Pillepich}, {Calistro Rivera}, {Schulz}, {Hernquist}, {Kaasinen}, {Marinacci}, {Nelson}, \& {Vogelsberger}}]{2022MNRAS.510.3321P}
{Popping}, G., {Pillepich}, A., {Calistro Rivera}, G., {et~al.} 2022, \mnras, 510, 3321, \dodoi{10.1093/mnras/stab3312}

\bibitem[{{Riechers} {et~al.}(2013){Riechers}, {Bradford}, {Clements}, {Dowell}, {P{\'e}rez-Fournon}, {Ivison}, {Bridge}, {Conley}, {Fu}, {Vieira}, {Wardlow}, {Calanog}, {Cooray}, {Hurley}, {Neri}, {Kamenetzky}, {Aguirre}, {Altieri}, {Arumugam}, {Benford}, {B{\'e}thermin}, {Bock}, {Burgarella}, {Cabrera-Lavers}, {Chapman}, {Cox}, {Dunlop}, {Earle}, {Farrah}, {Ferrero}, {Franceschini}, {Gavazzi}, {Glenn}, {Solares}, {Gurwell}, {Halpern}, {Hatziminaoglou}, {Hyde}, {Ibar}, {Kov{\'a}cs}, {Krips}, {Lupu}, {Maloney}, {Martinez-Navajas}, {Matsuhara}, {Murphy}, {Naylor}, {Nguyen}, {Oliver}, {Omont}, {Page}, {Petitpas}, {Rangwala}, {Roseboom}, {Scott}, {Smith}, {Staguhn}, {Streblyanska}, {Thomson}, {Valtchanov}, {Viero}, {Wang}, {Zemcov}, \& {Zmuidzinas}}]{2013Natur.496..329R}
{Riechers}, D.~A., {Bradford}, C.~M., {Clements}, D.~L., {et~al.} 2013, \nat, 496, 329, \dodoi{10.1038/nature12050}

\bibitem[{{Rizzo} {et~al.}(2023){Rizzo}, {Roman-Oliveira}, {Fraternali}, {Frickmann}, {Valentino}, {Brammer}, {Zanella}, {Kokorev}, {Popping}, {Whitaker}, {Kohandel}, {Magdis}, {Di Mascolo}, {Ikeda}, {Jin}, \& {Toft}}]{2023A&A...679A.129R}
{Rizzo}, F., {Roman-Oliveira}, F., {Fraternali}, F., {et~al.} 2023, \aap, 679, A129, \dodoi{10.1051/0004-6361/202346444}

\bibitem[{{Scoville} {et~al.}(2016){Scoville}, {Sheth}, {Aussel}, {Vanden Bout}, {Capak}, {Bongiorno}, {Casey}, {Murchikova}, {Koda}, {{\'A}lvarez-M{\'a}rquez}, {Lee}, {Laigle}, {McCracken}, {Ilbert}, {Pope}, {Sanders}, {Chu}, {Toft}, {Ivison}, \& {Manohar}}]{2016ApJ...820...83S}
{Scoville}, N., {Sheth}, K., {Aussel}, H., {et~al.} 2016, \apj, 820, 83, \dodoi{10.3847/0004-637X/820/2/83}

\bibitem[{{Sersic}(1968)}]{1968adga.book.....S}
{Sersic}, J.~L. 1968, {Atlas de Galaxias Australes}

\bibitem[{{Simpson} {et~al.}(2014){Simpson}, {Swinbank}, {Smail}, {Alexander}, {Brandt}, {Bertoldi}, {de Breuck}, {Chapman}, {Coppin}, {da Cunha}, {Danielson}, {Dannerbauer}, {Greve}, {Hodge}, {Ivison}, {Karim}, {Knudsen}, {Poggianti}, {Schinnerer}, {Thomson}, {Walter}, {Wardlow}, {Wei{\ss}}, \& {van der Werf}}]{2014ApJ...788..125S}
{Simpson}, J.~M., {Swinbank}, A.~M., {Smail}, I., {et~al.} 2014, \apj, 788, 125, \dodoi{10.1088/0004-637X/788/2/125}

\bibitem[{{Simpson} {et~al.}(2015){Simpson}, {Smail}, {Swinbank}, {Almaini}, {Blain}, {Bremer}, {Chapman}, {Chen}, {Conselice}, {Coppin}, {Danielson}, {Dunlop}, {Edge}, {Farrah}, {Geach}, {Hartley}, {Ivison}, {Karim}, {Lani}, {Ma}, {Meijerink}, {Micha{\l}owski}, {Mortlock}, {Scott}, {Simpson}, {Spaans}, {Thomson}, {van Kampen}, \& {van der Werf}}]{2015ApJ...799...81S}
{Simpson}, J.~M., {Smail}, I., {Swinbank}, A.~M., {et~al.} 2015, \apj, 799, 81, \dodoi{10.1088/0004-637X/799/1/81}

\bibitem[{{Simpson} {et~al.}(2017){Simpson}, {Smail}, {Swinbank}, {Ivison}, {Dunlop}, {Geach}, {Almaini}, {Arumugam}, {Bremer}, {Chen}, {Conselice}, {Coppin}, {Farrah}, {Ibar}, {Hartley}, {Ma}, {Micha{\l}owski}, {Scott}, {Spaans}, {Thomson}, \& {van der Werf}}]{2017ApJ...839...58S}
---. 2017, \apj, 839, 58, \dodoi{10.3847/1538-4357/aa65d0}

\bibitem[{{Smail} {et~al.}(1997){Smail}, {Ivison}, \& {Blain}}]{1997ApJ...490L...5S}
{Smail}, I., {Ivison}, R.~J., \& {Blain}, A.~W. 1997, \apjl, 490, L5, \dodoi{10.1086/311017}

\bibitem[{{Smail} {et~al.}(2021){Smail}, {Dudzevi{\v{c}}i{\={u}}t{\.{e}}}, {Stach}, {Almaini}, {Birkin}, {Chapman}, {Chen}, {Geach}, {Gullberg}, {Hodge}, {Ikarashi}, {Ivison}, {Scott}, {Simpson}, {Swinbank}, {Thomson}, {Walter}, {Wardlow}, \& {van der Werf}}]{2021MNRAS.502.3426S}
{Smail}, I., {Dudzevi{\v{c}}i{\={u}}t{\.{e}}}, U., {Stach}, S.~M., {et~al.} 2021, \mnras, 502, 3426, \dodoi{10.1093/mnras/stab283}

\bibitem[{{Smail} {et~al.}(2023){Smail}, {Dudzevi{\v{c}}i{\={u}}t{\.{e}}}, {Gurwell}, {Fazio}, {Willner}, {Swinbank}, {Arumugam}, {Summers}, {Cohen}, {Jansen}, {Windhorst}, {Meena}, {Zitrin}, {Keel}, {Cheng}, {Coe}, {Conselice}, {D'Silva}, {Driver}, {Frye}, {Grogin}, {Koekemoer}, {Marshall}, {Nonino}, {Pirzkal}, {Robotham}, {Rutkowski}, {Ryan}, {Tompkins}, {Willmer}, {Yan}, {Broadhurst}, {Diego}, {Kamieneski}, \& {Yun}}]{2023ApJ...958...36S}
{Smail}, I., {Dudzevi{\v{c}}i{\={u}}t{\.{e}}}, U., {Gurwell}, M., {et~al.} 2023, \apj, 958, 36, \dodoi{10.3847/1538-4357/acf931}

\bibitem[{{Smith} {et~al.}(2012){Smith}, {Lucey}, \& {Carter}}]{2012MNRAS.426.2994S}
{Smith}, R.~J., {Lucey}, J.~R., \& {Carter}, D. 2012, \mnras, 426, 2994, \dodoi{10.1111/j.1365-2966.2012.21922.x}

\bibitem[{{Springel} {et~al.}(2005){Springel}, {White}, {Jenkins}, {Frenk}, {Yoshida}, {Gao}, {Navarro}, {Thacker}, {Croton}, {Helly}, {Peacock}, {Cole}, {Thomas}, {Couchman}, {Evrard}, {Colberg}, \& {Pearce}}]{2005Natur.435..629S}
{Springel}, V., {White}, S. D.~M., {Jenkins}, A., {et~al.} 2005, \nat, 435, 629, \dodoi{10.1038/nature03597}

\bibitem[{{Stach} {et~al.}(2019){Stach}, {Dudzevi{\v{c}}i{\={u}}t{\.{e}}}, {Smail}, {Swinbank}, {Geach}, {Simpson}, {An}, {Almaini}, {Arumugam}, {Blain}, {Chapman}, {Chen}, {Conselice}, {Cooke}, {Coppin}, {da Cunha}, {Dunlop}, {Farrah}, {Gullberg}, {Hodge}, {Ivison}, {Kocevski}, {Micha{\l}owski}, {Miyaji}, {Scott}, {Thomson}, {Wardlow}, {Weiss}, \& {van der Werf}}]{2019MNRAS.487.4648S}
{Stach}, S.~M., {Dudzevi{\v{c}}i{\={u}}t{\.{e}}}, U., {Smail}, I., {et~al.} 2019, \mnras, 487, 4648, \dodoi{10.1093/mnras/stz1536}

\bibitem[{{Steidel} {et~al.}(1998){Steidel}, {Adelberger}, {Dickinson}, {Giavalisco}, {Pettini}, \& {Kellogg}}]{1998ApJ...492..428S}
{Steidel}, C.~C., {Adelberger}, K.~L., {Dickinson}, M., {et~al.} 1998, \apj, 492, 428, \dodoi{10.1086/305073}

\bibitem[{{Stevens} {et~al.}(2003){Stevens}, {Ivison}, {Dunlop}, {Smail}, {Percival}, {Hughes}, {R{\"o}ttgering}, {van Breugel}, \& {Reuland}}]{2003Natur.425..264S}
{Stevens}, J.~A., {Ivison}, R.~J., {Dunlop}, J.~S., {et~al.} 2003, \nat, 425, 264, \dodoi{10.1038/nature01976}

\bibitem[{{Sun} {et~al.}(2024){Sun}, {Helton}, {Egami}, {Hainline}, {Rieke}, {Willmer}, {Eisenstein}, {Johnson}, {Rieke}, {Robertson}, {Tacchella}, {Alberts}, {Baker}, {Bhatawdekar}, {Boyett}, {Bunker}, {Charlot}, {Chen}, {Chevallard}, {Curtis-Lake}, {Danhaive}, {DeCoursey}, {Ji}, {Lyu}, {Maiolino}, {Rujopakarn}, {Sandles}, {Shivaei}, {{\"U}bler}, {Willott}, \& {Witstok}}]{2024ApJ...961...69S}
{Sun}, F., {Helton}, J.~M., {Egami}, E., {et~al.} 2024, \apj, 961, 69, \dodoi{10.3847/1538-4357/ad07e3}

\bibitem[{{Swinbank} {et~al.}(2010){Swinbank}, {Smail}, {Chapman}, {Borys}, {Alexander}, {Blain}, {Conselice}, {Hainline}, \& {Ivison}}]{2010MNRAS.405..234S}
{Swinbank}, A.~M., {Smail}, I., {Chapman}, S.~C., {et~al.} 2010, \mnras, 405, 234, \dodoi{10.1111/j.1365-2966.2010.16485.x}

\bibitem[{{Tadaki} {et~al.}(2018){Tadaki}, {Iono}, {Yun}, {Aretxaga}, {Hatsukade}, {Hughes}, {Ikarashi}, {Izumi}, {Kawabe}, {Kohno}, {Lee}, {Matsuda}, {Nakanishi}, {Saito}, {Tamura}, {Ueda}, {Umehata}, {Wilson}, {Michiyama}, {Ando}, \& {Kamieneski}}]{2018Natur.560..613T}
{Tadaki}, K., {Iono}, D., {Yun}, M.~S., {et~al.} 2018, \nat, 560, 613, \dodoi{10.1038/s41586-018-0443-1}

\bibitem[{{Tamura} {et~al.}(2009){Tamura}, {Kohno}, {Nakanishi}, {Hatsukade}, {Iono}, {Wilson}, {Yun}, {Takata}, {Matsuda}, {Tosaki}, {Ezawa}, {Perera}, {Scott}, {Austermann}, {Hughes}, {Aretxaga}, {Chung}, {Oshima}, {Yamaguchi}, {Tanaka}, \& {Kawabe}}]{2009Natur.459...61T}
{Tamura}, Y., {Kohno}, K., {Nakanishi}, K., {et~al.} 2009, \nat, 459, 61, \dodoi{10.1038/nature07947}

\bibitem[{{Tamura} {et~al.}(2010){Tamura}, {Iono}, {Wilner}, {Kajisawa}, {Uchimoto}, {Alexander}, {Chung}, {Ezawa}, {Hatsukade}, {Hayashino}, {Hughes}, {Ichikawa}, {Ikarashi}, {Kawabe}, {Kohno}, {Lehmer}, {Matsuda}, {Nakanishi}, {Takata}, {Wilson}, {Yamada}, \& {Yun}}]{2010ApJ...724.1270T}
{Tamura}, Y., {Iono}, D., {Wilner}, D.~J., {et~al.} 2010, \apj, 724, 1270, \dodoi{10.1088/0004-637X/724/2/1270}

\bibitem[{{Tan} {et~al.}(2024){Tan}, {Daddi}, {Magnelli}, {Correa}, {Bournaud}, {Adscheid}, {Zhang}, {Elbaz}, {G{\'o}mez-Guijarro}, {Kalita}, {Liu}, {Liu}, {Pety}, {Puglisi}, {Schinnerer}, {Silverman}, \& {Valentino}}]{2024Natur.636...69T}
{Tan}, Q.-H., {Daddi}, E., {Magnelli}, B., {et~al.} 2024, \nat, 636, 69, \dodoi{10.1038/s41586-024-08201-6}

\bibitem[{{Toft} {et~al.}(2014){Toft}, {Smol{\v{c}}i{\'c}}, {Magnelli}, {Karim}, {Zirm}, {Michalowski}, {Capak}, {Sheth}, {Schawinski}, {Krogager}, {Wuyts}, {Sanders}, {Man}, {Lutz}, {Staguhn}, {Berta}, {Mccracken}, {Krpan}, \& {Riechers}}]{2014ApJ...782...68T}
{Toft}, S., {Smol{\v{c}}i{\'c}}, V., {Magnelli}, B., {et~al.} 2014, \apj, 782, 68, \dodoi{10.1088/0004-637X/782/2/68}

\bibitem[{{Tsukui} {et~al.}(2024){Tsukui}, {Wisnioski}, {Bland-Hawthorn}, {Mai}, {Iguchi}, {Baba}, \& {Freeman}}]{2024MNRAS.527.8941T}
{Tsukui}, T., {Wisnioski}, E., {Bland-Hawthorn}, J., {et~al.} 2024, \mnras, 527, 8941, \dodoi{10.1093/mnras/stad3588}

\bibitem[{{Umehata} {et~al.}(2014){Umehata}, {Tamura}, {Kohno}, {Hatsukade}, {Scott}, {Kubo}, {Yamada}, {Ivison}, {Cybulski}, {Aretxaga}, {Austermann}, {Hughes}, {Ezawa}, {Hayashino}, {Ikarashi}, {Iono}, {Kawabe}, {Matsuda}, {Matsuo}, {Nakanishi}, {Oshima}, {Perera}, {Takata}, {Wilson}, \& {Yun}}]{2014MNRAS.440.3462U}
{Umehata}, H., {Tamura}, Y., {Kohno}, K., {et~al.} 2014, \mnras, 440, 3462, \dodoi{10.1093/mnras/stu447}

\bibitem[{{Umehata} {et~al.}(2015){Umehata}, {Tamura}, {Kohno}, {Ivison}, {Alexander}, {Geach}, {Hatsukade}, {Hughes}, {Ikarashi}, {Kato}, {Izumi}, {Kawabe}, {Kubo}, {Lee}, {Lehmer}, {Makiya}, {Matsuda}, {Nakanishi}, {Saito}, {Smail}, {Yamada}, {Yamaguchi}, \& {Yun}}]{2015ApJ...815L...8U}
---. 2015, \apjl, 815, L8, \dodoi{10.1088/2041-8205/815/1/L8}

\bibitem[{{Umehata} {et~al.}(2017){Umehata}, {Tamura}, {Kohno}, {Ivison}, {Smail}, {Hatsukade}, {Nakanishi}, {Kato}, {Ikarashi}, {Matsuda}, {Fujimoto}, {Iono}, {Lee}, {Steidel}, {Saito}, {Alexander}, {Yun}, \& {Kubo}}]{2017ApJ...835...98U}
---. 2017, \apj, 835, 98, \dodoi{10.3847/1538-4357/835/1/98}

\bibitem[{{Umehata} {et~al.}(2018){Umehata}, {Hatsukade}, {Smail}, {Alexander}, {Ivison}, {Matsuda}, {Tamura}, {Kohno}, {Kato}, {Hayatsu}, {Kubo}, \& {Ikarashi}}]{2018PASJ...70...65U}
{Umehata}, H., {Hatsukade}, B., {Smail}, I., {et~al.} 2018, \pasj, 70, 65, \dodoi{10.1093/pasj/psy065}

\bibitem[{{Umehata} {et~al.}(2019){Umehata}, {Fumagalli}, {Smail}, {Matsuda}, {Swinbank}, {Cantalupo}, {Sykes}, {Ivison}, {Steidel}, {Shapley}, {Vernet}, {Yamada}, {Tamura}, {Kubo}, {Nakanishi}, {Kajisawa}, {Hatsukade}, \& {Kohno}}]{2019Sci...366...97U}
{Umehata}, H., {Fumagalli}, M., {Smail}, I., {et~al.} 2019, Science, 366, 97, \dodoi{10.1126/science.aaw5949}

\bibitem[{{Umehata} {et~al.}(2020){Umehata}, {Smail}, {Swinbank}, {Kohno}, {Tamura}, {Wang}, {Ao}, {Hatsukade}, {Kubo}, {Nakanishi}, \& {Hayatsu}}]{2020A&A...640L...8U}
{Umehata}, H., {Smail}, I., {Swinbank}, A.~M., {et~al.} 2020, \aap, 640, L8, \dodoi{10.1051/0004-6361/202038146}

\bibitem[{{Umehata} {et~al.}(2024){Umehata}, {Steidel}, {Smail}, {Swinbank}, {Monson}, {Rosario}, {Lehmer}, {Nakanishi}, {Kubo}, {Iono}, {Alexander}, {Kohno}, {Tamura}, {Ivison}, {Saito}, {Mitsuhashi}, {Huang}, \& {Matsuda}}]{2024arXiv241022155U}
{Umehata}, H., {Steidel}, C.~C., {Smail}, I., {et~al.} 2024, arXiv e-prints, arXiv:2410.22155, \dodoi{10.48550/arXiv.2410.22155}

\bibitem[{{Virtanen} {et~al.}(2020){Virtanen}, {Gommers}, {Oliphant}, {Haberland}, {Reddy}, {Cournapeau}, {Burovski}, {Peterson}, {Weckesser}, {Bright}, {van der Walt}, {Brett}, {Wilson}, {Millman}, {Mayorov}, {Nelson}, {Jones}, {Kern}, {Larson}, {Carey}, {Polat}, {Feng}, {Moore}, {VanderPlas}, {Laxalde}, {Perktold}, {Cimrman}, {Henriksen}, {Quintero}, {Harris}, {Archibald}, {Ribeiro}, {Pedregosa}, {van Mulbregt}, \& {SciPy 1. 0 Contributors}}]{2020NatMe..17..261V}
{Virtanen}, P., {Gommers}, R., {Oliphant}, T.~E., {et~al.} 2020, Nature Methods, 17, 261, \dodoi{10.1038/s41592-019-0686-2}

\bibitem[{{Walter} {et~al.}(2012){Walter}, {Decarli}, {Carilli}, {Bertoldi}, {Cox}, {da Cunha}, {Daddi}, {Dickinson}, {Downes}, {Elbaz}, {Ellis}, {Hodge}, {Neri}, {Riechers}, {Weiss}, {Bell}, {Dannerbauer}, {Krips}, {Krumholz}, {Lentati}, {Maiolino}, {Menten}, {Rix}, {Robertson}, {Spinrad}, {Stark}, \& {Stern}}]{2012Natur.486..233W}
{Walter}, F., {Decarli}, R., {Carilli}, C., {et~al.} 2012, \nat, 486, 233, \dodoi{10.1038/nature11073}

\bibitem[{{Wardlow} {et~al.}(2011){Wardlow}, {Smail}, {Coppin}, {Alexander}, {Brandt}, {Danielson}, {Luo}, {Swinbank}, {Walter}, {Wei{\ss}}, {Xue}, {Zibetti}, {Bertoldi}, {Biggs}, {Chapman}, {Dannerbauer}, {Dunlop}, {Gawiser}, {Ivison}, {Knudsen}, {Kov{\'a}cs}, {Lacey}, {Menten}, {Padilla}, {Rix}, \& {van der Werf}}]{2011MNRAS.415.1479W}
{Wardlow}, J.~L., {Smail}, I., {Coppin}, K.~E.~K., {et~al.} 2011, \mnras, 415, 1479, \dodoi{10.1111/j.1365-2966.2011.18795.x}

\bibitem[{{Wei{\ss}} {et~al.}(2013){Wei{\ss}}, {De Breuck}, {Marrone}, {Vieira}, {Aguirre}, {Aird}, {Aravena}, {Ashby}, {Bayliss}, {Benson}, {B{\'e}thermin}, {Biggs}, {Bleem}, {Bock}, {Bothwell}, {Bradford}, {Brodwin}, {Carlstrom}, {Chang}, {Chapman}, {Crawford}, {Crites}, {de Haan}, {Dobbs}, {Downes}, {Fassnacht}, {George}, {Gladders}, {Gonzalez}, {Greve}, {Halverson}, {Hezaveh}, {High}, {Holder}, {Holzapfel}, {Hoover}, {Hrubes}, {Husband}, {Keisler}, {Lee}, {Leitch}, {Lueker}, {Luong-Van}, {Malkan}, {McIntyre}, {McMahon}, {Mehl}, {Menten}, {Meyer}, {Murphy}, {Padin}, {Plagge}, {Reichardt}, {Rest}, {Rosenman}, {Ruel}, {Ruhl}, {Schaffer}, {Shirokoff}, {Spilker}, {Stalder}, {Staniszewski}, {Stark}, {Story}, {Vanderlinde}, {Welikala}, \& {Williamson}}]{2013ApJ...767...88W}
{Wei{\ss}}, A., {De Breuck}, C., {Marrone}, D.~P., {et~al.} 2013, \apj, 767, 88, \dodoi{10.1088/0004-637X/767/1/88}

\end{thebibliography}
\bibliographystyle{aasjournal}



\end{document}